\documentclass[cernpreprint, atlasdraft=false, LANGEDIT=false, LANGSHOW=false, texlive=2023, USenglish, texmf, orcidlogo]{atlasdoc}
 
\usepackage[subcaption=true, full]{atlaspackage}
\usepackage{atlasbiblatex}
 
\usepackage[process,jetetmiss]{atlasphysics}
 
\usepackage{multirow}
 
\graphicspath{{logos/}{./figures/}}
 
\usepackage{ANA-HIGG-2021-08-PAPER-defs}

\AtlasTitle{Constraints on Higgs boson production with large transverse momentum using $H\rightarrow b\bar{b}$ decays in the ATLAS detector}
 
\AtlasVersion{4.0} 
 
\PreprintIdNumber{CERN-EP-2021-185}
 
\AtlasJournal{Phys.\ Rev.\ D}
 
\AtlasJournalRef{Phys. Rev. D 105 (2022) 092003}
\AtlasDOI{10.1103/PhysRevD.105.092003}

\AtlasAbstract{
This paper reports constraints on Higgs boson production with transverse momentum above 1~TeV.
The analyzed data from proton--proton collisions at a center-of-mass energy of 13~TeV were recorded with the ATLAS detector at the Large Hadron Collider from 2015 to 2018 and correspond to an integrated luminosity of 136~fb$^{-1}$.
Higgs bosons decaying into $b\bar{b}$ are reconstructed as single large-radius jets recoiling against a hadronic system and identified by the experimental signature of two $b$-hadron decays.
The experimental techniques are validated in the same kinematic regime using the $Z\rightarrow b\bar{b}$ process.
The 95\% confidence-level upper limit on the cross section for Higgs boson production with transverse momentum above 450~GeV is 115~fb, and
above 1~TeV it is 9.6~fb.
The Standard Model cross section predictions for a Higgs boson with a mass of 125~GeV
in the same kinematic regions
are 18.4~fb and 0.13~fb, respectively.
}
 
\author{The ATLAS Collaboration}
 
\AtlasRefCode{HIGG-2021-08}

\hypersetup{pdftitle={ATLAS document},pdfauthor={The ATLAS Collaboration}}
 
\begin{document}
 
\maketitle

\section{Introduction}

The characterization of the Higgs sector has steadily improved since the Higgs boson (\Hboson) discovery~\cite{HIGG-2012-27,CMS-HIG-12-028} using data from proton--proton ($pp$) collisions produced by the Large Hadron Collider (LHC) at CERN.
Four production modes have been observed,
gluon--gluon fusion (\ggF),
vector-boson fusion (\VBF),
associated production with a weak vector boson (\VH), and
associated production with a top quark--antiquark pair (\ttH),
along with five decay modes
$H \to \gamma\gamma$,
$ZZ^\ast$,
$WW^\ast$,
$\tau\tau$,
$b\bar{b}$~\cite{HIGG-2018-57, CMS-HIG-17-031}.
The initial measurements of inclusive cross sections have evolved to include differential cross-section measurements, and measurements in the simplified template cross-section framework~\cite{deFlorian:2016spz,Badger:2016bpw,Amoroso:2020lgh}.
All results agree with the Standard Model (SM) predictions within the current precision, but
sizable regions of the Higgs sector remain unexplored.
In one such region, where the Higgs boson transverse momentum \pTH\ reaches the \TeV\ scale, the cross-section hierarchy is very different from that in the inclusive cross section, where \ggF\ is nearly 90\% of the total.
At the \TeV\ scale, the SM predicts the cross sections for the \ggF\ and \VH\ production processes to be roughly equal, while the \VBF\ and \ttH\ production cross sections are around 60\% and 30\% of the \ggF\ process, respectively.
 
The leading effects of many beyond-the-SM (BSM) scenarios can be parameterized through effective field theories (EFTs), whose operators are suppressed by a new physics scale $\Lambda$~\cite{Contino:2013kra}.
Measured observables at the LHC would only be affected through effective interactions among SM particles.
For example, the \ggF\ production mode is sensitive to the structure of quasi-point-like couplings within the loop nature of the effective $ggH$ coupling.
Studies of Higgs bosons produced with large transverse momentum access regions where some potential BSM effects are enhanced by powers of $\pTH/\Lambda$~\cite{Grojean:2013nya,Biekoetter:2014jwa,Mimasu:2015nqa,Grazzini:2016paz,Banfi:2018pki}. Differential cross-section measurements with an extended reach may be more sensitive than higher precision, low-energy measurements also because the signal-to-background ratio increases with \pTH.
 
In the high-\pTH regime, the CMS Collaboration measured a signal yield relative to the SM prediction, or signal strength, of $\mu_{\ggF}=3.7^{+1.6}_{-1.5}$ in the \Hbb decay mode for events containing a large-radius jet with $\pT>450~\GeV$, and presented \ggF\ differential cross sections while considering other Higgs boson production modes as a background~\cite{CMS-HIG-19-003}.
In the \Hgg\ decay mode, \ggF\ production with $\pTH>200~\GeV$ was measured to a precision of less than 50\% relative to the SM prediction~\cite{CMS-HIG-19-015}.
The analysis of \VH\ production with leptonic \Vboson\ decays has achieved considerable sensitivity in the high-\pTH\ regime~\cite{HIGG-2018-51,HIGG-2018-52,CMS-HIG-18-016}.
The ATLAS Collaboration measured a signal strength of $\mu_{\VH}=0.72^{+0.39}_{-0.36}$ in the \Hbb decay mode targeting events with $\pTH>250~\GeV$, and presented differential cross sections in two exclusive vector-boson transverse momentum regions, 250--400~\GeV\ and above 400~\GeV~\cite{HIGG-2018-52}.
However, it is sensitive to different EFT operators than those for \ggF\ and \ttH\ production.
Recent results on \VBF\ production with Higgs boson decays to photons and leptons also included high-\pTH\ event categories, but have limited reach~\cite{HIGG-2017-07,HIGG-2018-28,HIGG-2019-04A,CMS-HIG-19-001,CMS-HIG-14-004,}. Similarly, measurements of \ttH\ production have yet to reach the high-\pTH\ regime~\cite{HIGG-2018-13,CMS-HIG-19-008}.
 
This paper reports the first ATLAS studies of Higgs bosons produced with transverse momentum above 1~\TeV.
The yield of Higgs bosons decaying into \bbbar\ pairs is determined in several \pTH regions.
No restrictions are applied to select a particular Higgs boson production mode aside from requiring an energetic hadronic recoil system.
The cross section is measured for $\pTH>450$~\GeV, allowing a straightforward comparison with theoretical calculations, such as those reported in Ref.~\cite{Becker:2020rjp}.
In addition a differential analysis is performed to extract the cross section in four Higgs boson \pT intervals, 300--450~\GeV, 450--650~\GeV, 650--1000~\GeV, and above 1~\TeV.
 
Data used correspond to 136~\ifb\ of $pp$ collisions at $\sqrt{s}=13~\TeV$ and were collected with the ATLAS detector~\cite{PERF-2007-01} using jet-based trigger requirements during Run~2 (2015--2018) of the LHC.
Higgs bosons with a large Lorentz boost are reconstructed as single large-radius jets having a mass compatible with 125~\GeV~\cite{BDRS}.
To populate the signal region, events must have at least two large-radius jets.
At least one jet must have $\pT>450~\GeV$ to ensure a fully efficient trigger response.
Either of the two leading-\pT jets in the event must contain evidence of two $b$-hadron decays.
Including the subleading jet as a possible Higgs boson candidate increases the sensitivity for $\pTH>450~\GeV$ by 11\% and permits a cross-section measurement down to $\pTH=300~\GeV$, overlapping with measurements in other decay channels~\cite{HIGG-2016-21,HIGG-2018-13,HIGG-2018-28,HIGG-2018-29,CMS-HIG-19-015,CMS-HIG-19-001,CMS-HIG-19-008A}.
 
For a Higgs boson mass (\mH) of $125~\GeV$, the SM predicts the \ggF\ production mode contributes nearly half of the Higgs boson events reconstructed near \mH when they are summed over all signal regions. \VBF-, \VH-, and \ttH-produced events each contribute approximately 15\%--20\%.
Since some of the same dimension-6 EFT operators modify the \pTH spectrum in the \ggF\ and \ttH\ processes, it could be advantageous to consider \ttH\ and \ggF\ production together.
These operators can produce enhancements at high \pTH\ that are within the sensitivity of the present analysis without inducing significant deviations from the SM prediction at low \pTH.
However, the expected yield enhancement differs for the two production modes, with \ggF\ production having a larger increase.
While this analysis is primarily sensitive to \ggF\ production, all the main production modes are considered as the signal.
This approach enhances the sensitivity to possible BSM effects and minimizes the dependence on theoretical assumptions.
Using the \Hbb decay, which has the largest branching fraction, mitigates the impact of the smaller absolute cross section in the high-\pT regime.
 
The dominant background process is multijet production, which exhibits a monotonically decreasing jet mass distribution.
Hadronically decaying vector bosons, produced in association with jets (\Vjets) and events with one or two top quarks (jointly referred to as \textit{\T}) populate the jet mass regions below and above \mH, respectively, as shown in Figure~\ref{fig:resShapes}.
The \Zboson and \Hboson resonance structures are distinct from the smoothly falling multijet background, while the top quark resonance is spread over a large portion of the high-mass region.
Therefore, the signal is extracted from the reconstructed jet mass distribution.
An analytic function is used to model the multijet background.
The jet mass spectra and acceptance for Higgs boson, \Vjets, and \T\ events are estimated from simulation.
 
A binned maximum-likelihood fit, referred to as the global likelihood fit, is used to measure the signal strength.
Unconstrained, or free, parameters for the multijet model's yield and shape, the \Zjets\ yield, and the Higgs boson yield are determined simultaneously from the signal region data.
Control region data, fit simultaneously with the signal region, determine the unconstrained yield for top quark pair-production (\ttbar) events.
Concurrently, the modeling of the jet mass scale and resolution from \Vboson\ and top quark decays are separately validated in a broad range of jet \pT.
Validation region data are used to study \Vjets\ events, the parameterization of the multijet model, and the robustness of the global likelihood fit.

\begin{figure}[htb]
\begin{center}
\begin{subfigure}{0.45\textwidth}
\centering
\includegraphics[width=\textwidth]{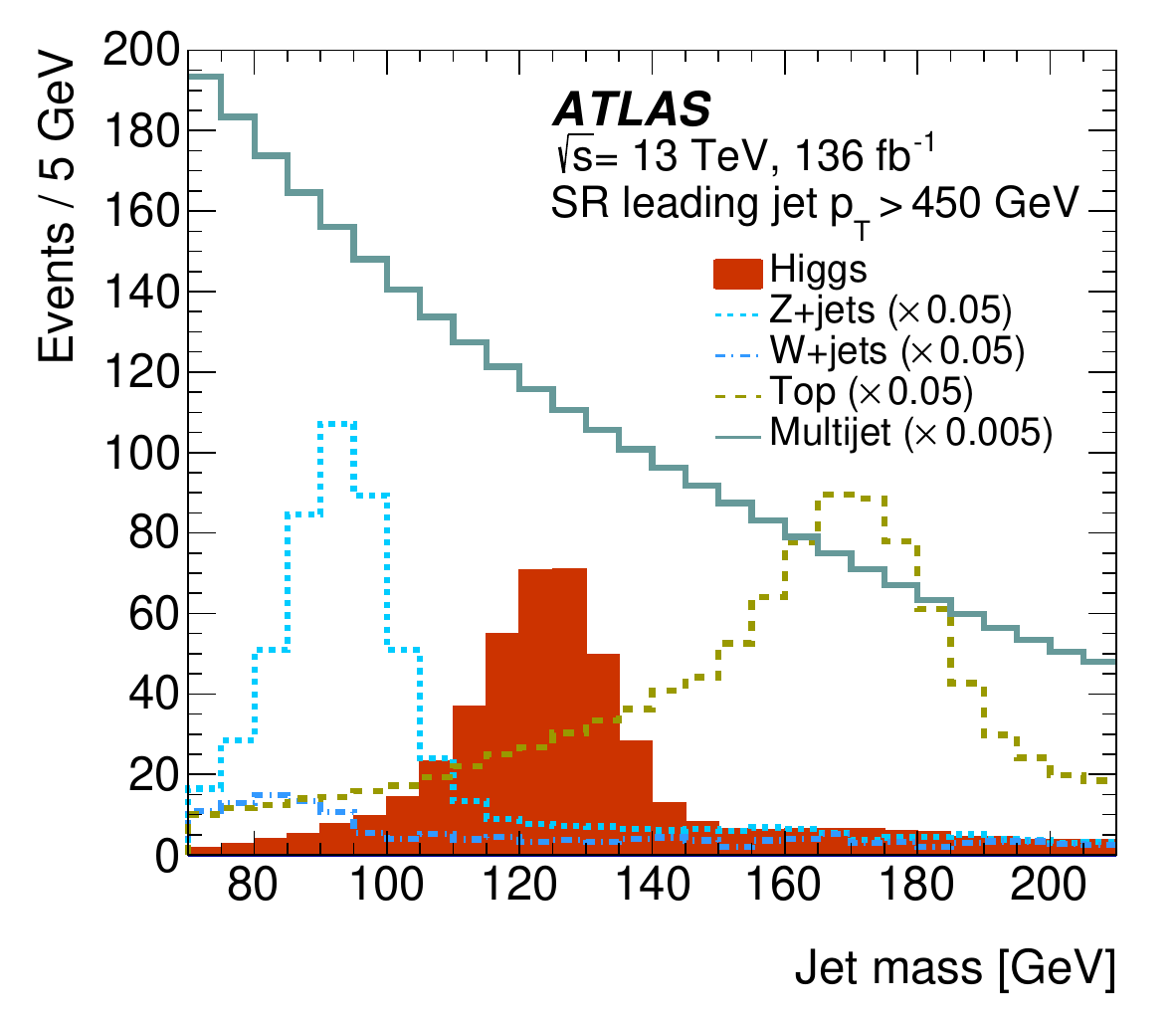}
\end{subfigure}
\begin{subfigure}{0.45\textwidth}
\centering
\includegraphics[width=\textwidth]{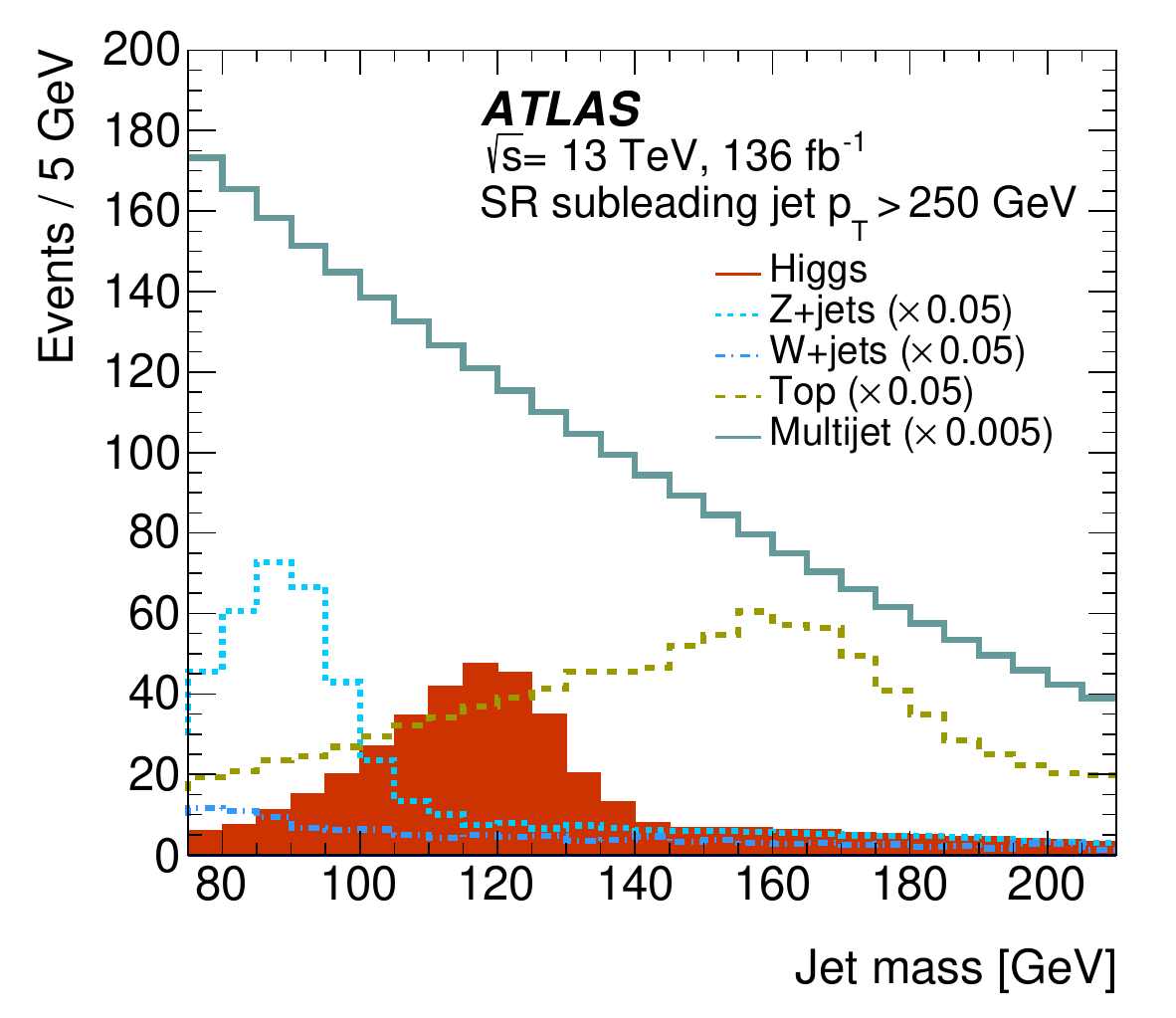}
\end{subfigure}
\end{center}
\caption{Jet mass distributions for the Higgs boson, \Zjets, \Wjets, and \T\ contributions from the SM prediction as well as the multijet jet mass distribution extracted from data in the signal region (SR) defined by the leading (left) and subleading (right) jets. 
\label{fig:resShapes}}
\end{figure}


\section{ATLAS detector}

\newcommand{\AtlasCoordFootnote}{
ATLAS uses a right-handed coordinate system with its origin at the nominal interaction point (IP)
in the center of the detector and the \(z\)-axis along the beam pipe.
The \(x\)-axis points from the IP to the center of the LHC ring,
and the \(y\)-axis points upwards.
Cylindrical coordinates \((r,\phi)\) are used in the transverse plane,
\(\phi\) being the azimuthal angle around the \(z\)-axis.
The pseudorapidity is defined in terms of the polar angle \(\theta\) as \(\eta = -\ln \tan(\theta/2)\).
Angular distance is measured in units of \(\Delta R \equiv \sqrt{(\Delta\eta)^{2} + (\Delta\phi)^{2}}\).}
 
The ATLAS experiment~\cite{PERF-2007-01} at the LHC is a multipurpose particle detector
with a forward--backward symmetric cylindrical geometry and a near \(4\pi\) coverage in
solid angle.\footnote{\AtlasCoordFootnote}
It consists of an inner tracking detector surrounded by a thin superconducting solenoid
providing a \SI{2}{\tesla} axial magnetic field, electromagnetic and hadron calorimeters, and a muon spectrometer.
The inner tracking detector covers the pseudorapidity range \(|\eta| < 2.5\).
It consists of silicon pixel, silicon microstrip, and transition radiation tracking detectors.
The Insertable B-Layer, the innermost pixel layer at a mean radius of 3.3~cm, was installed before Run~2 of the LHC~\cite{ATLAS-TDR-2010-19,PIX-2018-001}.
Lead/liquid-argon (LAr) sampling calorimeters provide electromagnetic (EM) energy measurements
with high granularity.
A steel/scintillator-tile hadron calorimeter covers the central pseudorapidity range (\(|\eta| < 1.7\)).
The forward regions are instrumented with LAr calorimeters
for EM and hadronic energy measurements up to \(|\eta| = 4.9\).
The muon spectrometer surrounds the calorimeters and is based on
three large air-core toroidal superconducting magnets with eight coils each.
The field integral of the toroids ranges between \num{2.0} and \SI{6.0}{\tesla\metre}
across most of the detector.
The muon spectrometer includes a system of precision chambers for tracking and fast detectors for triggering.
A two-level trigger system is used to select events~\cite{TRIG-2019-04}.
The first-level trigger is implemented in hardware and uses a subset of the detector information to accept events at a rate below \SI{100}{\kHz}.
A software-based trigger reduces the accepted event rate to \SI{1}{\kHz} on average. 
An extensive software suite~\cite{ATL-SOFT-PUB-2021-001} is used in the reconstruction and analysis of real and simulated data,
in detector operations, and in the trigger and data acquisition systems of the experiment.


\section{Data and simulated samples}
\label{sec:samples}

The data were collected with the ATLAS detector in $pp$ collisions with a center-of-mass energy of 13~\TeV\ during Run 2 (2015--2018) of the LHC~\cite{TRIG-2016-01}.
Events must satisfy a set of triggers requiring a jet reconstructed with the \antikt\ algorithm with radius parameter $R=1.0$~\cite{Cacciari:2008gp}.
To adapt to different instantaneous luminosity profiles and the inclusion of pileup suppression techniques within the data acquisition system~\cite{PERF-2014-03}, the jet $\pT$ and mass thresholds differ for each year of data taking.
The trigger jet-$\pT$ threshold varies from $360~\GeV$ to $420~\GeV$, and the trigger jet-mass threshold is either not applied, $30~\GeV$, or $35~\GeV$.
Events which pass a trigger requiring a muon with $\pT>50$~GeV~\cite{TRIG-2018-01} populate a control region for top quark production.
The total integrated luminosities are 136~\ifb\ and 139~\ifb\ for
the jet- and muon-triggered data, respectively,
with an uncertainty of 1.7\%~\cite{ATLAS-CONF-2019-021,LUCID2}.
 
Monte Carlo (MC) simulated events are used to model the resonant backgrounds
(\Wjets, \Zjets, and \T\ production) as well as four Higgs production processes:
\ggF, \VBF, \VH, and \ttH.
Higgs boson \ggF\ production was simulated at next-to-leading-order (NLO) accuracy in QCD with finite mass effects by using the {\textsc Hj-MiNLO}~\cite{Hamilton:2012np,Campbell:2012am,Hamilton:2012rf} prescription
with the \textsc{Powheg} program~\cite{Alioli:2010xd,Nason:2004rx,Frixione:2007vw} as discussed in Ref.~\cite{Hamilton:2015nsa}.
NLO accuracy in QCD for \VBF\ and \ttH\ production and LO accuracy for $gg\to\VH$ production was achieved using the \POWHEGBOX[v2]~\cite{Nason:2009ai,Hartanto:2015uka,Alioli:2010xd,Nason:2004rx,Frixione:2007vw} program.
Using the \POWHEGBOX[v2] program, the improved \minlo~\cite{Luisoni:2013cuh} calculation, and the \gosam~\cite{Cullen:2011ac} program, $qq\to\VH$ production was also simulated at NLO accuracy in QCD.
Corrections for NLO electroweak (EW) effects
were applied as a function of the generated Higgs boson transverse momentum for \VBF, \VH, and \ttH\ production.
The production cross sections used are compatible with those presented in Ref.~\cite{Becker:2020rjp}, except for \ttH\ production where a scale factor is applied to match the reported value~\cite{Becker:2020rjp}.
The Higgs boson
branching fractions were calculated with \textsc{Hdecay}~\cite{Djouadi:1997yw,Spira:1997dg,Djouadi:2006bz} and \PROPHECY~\cite{Bredenstein:2006ha,Bredenstein:2006rh,Bredenstein:2006nk}.

Production of \Vjets\ events with hadronic boson decays was simulated with \SHERPA to NLO QCD accuracy for one additional parton and LO QCD accuracy for up to four additional partons. Approximate NLO EW corrections~\cite{Bothmann:2019yzt} were applied as a function of the generated vector-boson momentum \VpT. They have a sizable impact on the differential production cross section, reducing the predicted yield by ${\sim}10$\% at a \VpT of 500~\GeV\ and ${\sim}20$\% above 1~\TeV.
Calculations of next-to-next-to-leading-order (NNLO) QCD corrections to \Vjets production are available~\cite{Blumenschein:2018gtm}. The NNLOJET group performed the calculation for $\sqrt{s}=8~\TeV$~\cite{Ridder:2015dxa,Ridder:2016nkl} and has provided custom corrections for the analysis kinematic region for $\sqrt{s}=13~\TeV$ as a function \VpT. They vary from 1.01 to 1.09 and are applied as a multiplicative factor on top of the NLO EW corrections.
Diboson production was found to make a negligible contribution to the present analysis.

\begin{table}[tb]
\begin{center}{\fontsize{7}{7.9}\selectfont
\caption{The generators used for the simulation of the signal and background processes.
Matrix element, parton shower and underlying event are abbreviated as ME, PS, and UE respectively.
$(\ast)$ \POWHEG was configured to output events with Born $k_\mathrm{T}$ above 200~$\GeV$ using the {\em bornktmin} setting.
$(\bullet)$ Corrections for NLO EW effects computed with \textsc{Hawk}~\cite{Denner:2011id,Denner:2014cla} are applied as a function of the generated Higgs boson transverse momentum. 
$(\circ)$ Corrections for NLO EW effects computed with \textsc{Sherpa}$+$\openloops~\cite{Buccioni:2019sur,Cascioli:2011va,Denner:2016kdg} are applied as a function of the generated Higgs boson transverse momentum and were provided by Ref.~\cite{Becker:2020rjp}. 
$(\dagger)$ \SHERPA provides one additional parton at NLO accuracy and up to four additional partons at LO in QCD and custom NNLO QCD corrections were provided by the NNLOJET group.
\protect\label{tab:samples}}
\begin{tabular}{lccccc}
\toprule
\hspace*{0.3cm} Process & ME generator & ME PDF &PS and & UE model & Cross-section \hspace{2.5cm}\\
& & & hadronization & tune & order\\
\midrule
\multicolumn{6}{l}{Higgs Boson} \\
\midrule
$gg\to H \to b\bar{b} $
& \parbox{0.17\textwidth}{\centering \POWHEGBOX[v2]$^{(\ast)}$~\cite{Alioli:2010xd,Nason:2004rx,Frixione:2007vw} + \minlo~\cite{Hamilton:2012np,Campbell:2012am,Hamilton:2012rf}}
& \NNPDF[3.0nnlo]~\cite{Ball:2014uwa}
& \PYTHIA[8.212]~\cite{Sjostrand:2014zea}
& AZNLO~\cite{STDM-2012-23}
& \parbox{0.12\textwidth}{\centering NLO(QCD) + LO(EW)} \\
$qq\to H \to q'q'b\bar{b} $
& \POWHEGBOX[v2]~\cite{Nason:2009ai,Alioli:2010xd,Nason:2004rx,Frixione:2007vw}
& \NNPDF[3.0nlo]~\cite{Ball:2014uwa}
& \PYTHIA[8.230]
& AZNLO
& \parbox{0.12\textwidth}{\centering NLO(QCD) + NLO(EW)$^{(\bullet)}$} \\
$qq\to WH$  \\
\hspace*{0.7cm} $\to qq' b\bar{b}$
& \multirow{2}{*}{\parbox{0.17\textwidth}{\centering\POWHEGBOX[v2] + \gosam~\cite{Cullen:2011ac} + \minlo~\cite{Luisoni:2013cuh}}}
& \multirow{2}{*}{\NNPDF[3.0nlo]}
& \PYTHIA[8.240]
& \multirow{2}{*}{AZNLO}
& \multirow{2}{*}{\parbox{0.12\textwidth}{\centering NNLO(QCD) + NLO(EW)$^{(\bullet)}$}} \\
\hspace*{0.7cm} $\to \ell\nu b\bar{b}$ & & & \PYTHIA[8.212] & & \\
$qq\to ZH$ \\
\hspace*{0.7cm}$\to q\bar{q}b\bar{b}$
& \multirow{3}{*}{\parbox{0.17\textwidth}{\centering\POWHEGBOX[v2] + \gosam\ + \minlo}}
& \multirow{3}{*}{\NNPDF[3.0nlo]}
& \PYTHIA[8.240]
& \multirow{3}{*}{AZNLO}
& \multirow{3}{*}{\parbox{0.12\textwidth}{\centering NNLO(QCD)
+ NLO(EW)$^{(\bullet)}$}} \\
\hspace*{0.7cm}$\to\nu\nu   b\bar{b}$ & & & \multirow{2}{*}{\PYTHIA[8.212]} & & \\
\hspace*{0.7cm}$\to\ell\ell b\bar{b}$ & & & & & \\
$gg\to ZH$ \\
\hspace*{0.7cm}$\to q\bar{q}b\bar{b}$
& \multirow{3}{*}{\POWHEGBOX[v2]~\cite{Hartanto:2015uka}}
& \multirow{3}{*}{\NNPDF[3.0nlo]}
& \PYTHIA[8.240]
& \multirow{3}{*}{AZNLO}
& \multirow{3}{*}{\parbox{0.12\textwidth}{\centering LO + NLL(QCD)}} \\
\hspace*{0.7cm}$\to\nu\nu   b\bar{b}$ & & & \multirow{2}{*}{\PYTHIA[8.212]} & & \\
\hspace*{0.7cm}$\to\ell\ell b\bar{b}$ & & & & & \\
$gg\to ttH$ \\
\hspace*{0.7cm}$tt\to\text{all}$
& \multirow{2}{*}{\POWHEGBOX[v2]}
& \multirow{2}{*}{\NNPDF[3.0nlo]}
& \multirow{2}{*}{\PYTHIA[8.230]}
& \multirow{2}{*}{AZNLO}
& \multirow{2}{*}{\parbox{0.12\textwidth}{\centering NLO(QCD) + NLO(EW)$^{(\circ)}$}} \\
\hspace*{0.7cm}$H\to\text{all}$  & & & & & \\
\midrule
\multicolumn{6}{l}{Vector boson + jets} \\
\midrule
$W\to qq$ & \multirow{2}{*}{\SHERPA[2.2.8]~\cite{Gleisberg:2008ta,Cascioli:2011va,Gleisberg:2008fv}} & \multirow{2}{*}{\NNPDF[3.0nnlo]} &  \multirow{2}{*}{\SHERPA[2.2.8]~\cite{Schumann:2007mg,Hoeche:2012yf}} & \multirow{2}{*}{Default} & NNLO(QCD)~$^{(\dagger)}$~\cite{Catani:2009sm,Ridder:2015dxa,Ridder:2016nkl} \\
$Z\to qq$ & & & & & approx NLO(EW)~\cite{Kallweit:2014xda,Biedermann:2017yoi,Bothmann:2019yzt} \\
\midrule
\multicolumn{6}{l}{Top quark, mass set to 172.5~\GeV}  \\
\midrule
$t\bar{t}\to\text{all}$
&\POWHEGBOX[v2]~\cite{Frixione:2007nw,Nason:2004rx,Frixione:2007vw,Alioli:2010xd} & \NNPDF[3.0nlo] &\PYTHIA[8.230] &A14~\cite{ATL-PHYS-PUB-2014-021}& NNLO+NNLL~\cite{Czakon:2011xx} \\
$tW$
&\POWHEGBOX[v2]~\cite{Re:2010bp,Nason:2004rx,Frixione:2007vw,Alioli:2010xd}       & \NNPDF[3.0nlo] & \PYTHIA[8.230] & A14 & NLO \\ 
$t$ t-channel
&\POWHEGBOX[v2]~\cite{Frederix:2012dh,Nason:2004rx,Frixione:2007vw,Alioli:2010xd} & \NNPDF[3.0nlo] & \PYTHIA[8.230] & A14 & NLO \\ 
$t$ s-channel
&\POWHEGBOX[v2]~\cite{Alioli:2009je,Nason:2004rx,Frixione:2007vw,Alioli:2010xd}   & \NNPDF[3.0nlo] & \PYTHIA[8.230] & A14 & NLO \\ 
\midrule
\multicolumn{6}{l}{Multijet}  \\
\midrule
Dijets
&\PYTHIA[8.230] &  \NNPDF[2.3lo]~\cite{Ball:2012cx} &\PYTHIA[8.230] &A14& LO \\
\bottomrule
\end{tabular}}
\end{center}
\end{table}


The production of top quark pairs, associated production of a top quark with a $W$ boson ($tW$), and single-top t- and s-channel production were modeled using the \POWHEGBOX[v2]~\cite{Frixione:2007nw,Re:2010bp,Frederix:2012dh,Alioli:2009je,Nason:2004rx,Frixione:2007vw,Alioli:2010xd} generator at NLO in QCD.
The diagram subtraction scheme~\cite{Frixione:2008yi} was used in $tW$ events to account for interference and overlap with \ttbar\ production.

The jet mass distribution of nonresonant multijet events is modeled with an analytic function.
Simulated events used to study the multijet model were generated using \PYTHIA[8.230]~\cite{Sjostrand:2014zea} with leading-order matrix elements for dijet production and interfaced to a \pt-ordered parton shower.

All simulated particles from collisions were processed with the ATLAS detector simulation~\cite{SOFT-2010-01} based on \GEANT~\cite{Agostinelli:2002hh}.
Pileup, multiple interactions in the same and neighboring bunch crossings, was modeled by overlaying the hard-scatter event with inelastic $pp$ events generated with \PYTHIA[8.186]~\cite{Sjostrand:2014zea} using the \NNPDF[2.3lo] set of parton distribution functions (PDFs)~\cite{Ball:2012cx} and a set of tuned parameters called the A3 tune~\cite{ATL-PHYS-PUB-2016-017}.
For Higgs boson, \T, and dijet production, the \EVTGEN[1.2.0] program~\cite{Lange:2001uf} was used to model the decays of bottom and charm hadrons.
 
For each sample, Table~\ref{tab:samples} summarizes the MC generator, parton distribution functions, parton shower and hadronization model, and underlying-event tune used, as well as the order of perturbative QCD computations and EW corrections obtained for the cross section.
For additional information, see
Ref.~\cite{ATL-PHYS-PUB-2017-006} for \Vjets events,
Refs.~\cite{ATL-PHYS-PUB-2016-020,ATL-PHYS-PUB-2016-004, ATL-PHYS-PUB-2017-007,ATL-PHYS-PUB-2018-009} for top quark events,
and
Ref.~\cite{ATL-PHYS-PUB-2019-017} for multijet events.
Systematic uncertainties for process modeling are described in Section~\ref{sec:background}.


\section{Object selection}

For Higgs bosons with a large Lorentz boost, the event topology of $pp\to H(\to b\bbar)+J$
is characterized by two jets, one of which contains the decay products of the two $b$-hadrons.
 
\subsection{Object reconstruction}
 
Charged-particle tracks~\cite{ATL-PHYS-PUB-2015-051} are reconstructed in the inner detector and
used to form interaction vertices~\cite{PERF-2015-01}.
The primary vertex of the hard interaction is defined as the vertex with the highest sum of squared transverse momenta of associated tracks.
 
Large-radius ($R=1.0$) jets are formed by applying the \antikt\ algorithm implemented in \Fastjet~\cite{Fastjet} to topological clusters of noise-suppressed calorimeter energy depositions calibrated to the local hadronic scale~\cite{PERF-2014-07}.
Jet cleaning criteria are applied to identify jets arising from noncollision backgrounds or noise in the calorimeters~\cite{ATLAS-CONF-2015-029}, and events containing such jets are removed.
A jet trimming procedure reduces pileup dependence and improves the mass resolution~\cite{Krohn:2009th}. It produces a collection of subjets by reclustering the constituents of each jet using the \kt\ algorithm~\cite{ktAlgo} with $R=0.2$. Subjets with $\pT^\text{subjet}/\pT^\text{jet} < 0.05$ are removed and the jet four-momentum is recalculated.
The trimmed jet's mass \mj is computed as a weighted combination of the jet mass obtained from the calorimeter measurements and that from the charged component in the inner detector~\cite{JETM-2018-02}.
For jets within $\abseta<2$, simulation-based corrections are applied to calibrate \pT\ and \mj,
while \pT also has corrections based on \insitu techniques~\cite{JETM-2018-02}.
 
The \antikt\ algorithm with a variable, \pT-dependent radius parameter is used to form \trkjets\ from tracks compatible with the primary vertex~\cite{Krohn:2009zg,ATL-PHYS-PUB-2017-010}. The effective jet radius $R_{\text{eff}}$ is $\rho/\pT$ where the $\rho$-parameter is set to 30~\GeV. The lower and upper bounds on the track-jet radius are 0.02 and 0.4, respectively.
\Trkjets\ are matched to large-$R$ jets by ghost association before trimming~\cite{Cacciari:2008gn,Cacciari:2007fd}.
In simulated events, \trkjets\ are labeled as $b$-, $c$- or light-flavor according to which hadrons with $\pt>5~\GeV$ are found within $\Delta R=0.3$ of the jet axis~\cite{FTAG-2018-01}.
 
A multivariate discriminant (denoted MV2 in Ref.~\cite{FTAG-2018-01}) is used to tag \trkjets\ containing a $b$-hadron decay (\btag).
\Trkjets\ with $\pT>10~\GeV$, $\abseta < 2.5$, and at least two tracks are considered.
The operating point is tuned to produce an average efficiency of 77\% for \bjets in simulated \ttbar\ events, which corresponds to light-flavor jet ($u$-, $d$-, $s$-quark, and gluon) and $c$-jet misidentification efficiencies of 0.9\% and 25\%, respectively.
If the $\Delta R$ between two \trkjets\ with $\pT>5~\GeV$ in a large-$R$ jet is smaller than either of their respective variable radii, the jet is not considered for \btag~\cite{PERF-2017-04}.
 
Muons are required to have $\abseta<2.5$, $p_{\text{T}}^{\mu}>10~\GeV$, and small impact parameters relative to the primary vertex,
as well as to satisfy the `medium' quality criterion~\cite{MUON-2018-03}.
Isolated muons must also satisfy loose track- and calorimeter-based isolation conditions~\cite{MUON-2018-03}.

\subsection{Analysis object definitions}
Reconstructed jets possessing properties compatible with an \Hbb decay are labeled \textit{candidate jets}.
The reconstructed jet containing the Higgs boson decay products, or \textit{H-jet}, is not always the highest-\pT\ jet in the event.
The Higgs boson and the hadronic recoil system have equal \pT, but the \pT\ ordering of the reconstructed jets is affected by final-state radiation, jet resolution, and activity outside the jet cone.
Undetected neutrinos from semileptonic $b$-hadron decays may also cause the \Hboson-jet to be reconstructed as the subleading jet.
In around 50\% (47\%) of selected simulated \ggF\ events, the \Hboson-jet is the leading (subleading) jet.
Simulated \VH\ events contain similar proportions of leading and subleading \Hboson-jets.
For \ttH\ production, any one of at least three final-state particles can be reconstructed as the leading jet.
Therefore, candidate jets are defined as \textit{either} of the two leading-\pT\ jets with $\abseta<2$,
$\pt>250~\GeV$, $\mj>60~\GeV$, and $2\mj/\pt<1$. The last requirement selects jets compatible with a boosted decay.
Furthermore, they must contain at least two \trkjets.
A candidate jet is \textit{double-tagged} if its two leading \trkjets\ are \btagged and \textit{anti-tagged} if neither are \btagged.
Each signal region requires a double-tagged candidate jet, as discussed in Section~\ref{sec:SRandVR}.
 
A `muon-in-jet' correction is applied to candidate jets to account for the presence of semileptonic $b$-hadron decays.
It utilizes the leading-\pT\ muon passing a minimum set of quality criteria and found within
$\Delta R=\min(0.4, 0.04 + 10/p_{\text{T}}^{\mu})$
of a \btagged \trkjet. 
The muon four-momentum is added to the trimmed jet while the energy deposited by the muon in the calorimeter is removed.
After correcting 13\% (33\%) of leading (subleading) $H$-jets in simulated \ggF\ events, the \mj\ width is reduced by 5\% (12\%).
Henceforth, \pT\ and \mJ\ refer to the muon-corrected jet transverse momentum
and mass, respectively, and \pTU\ and \mJU\ represent the corresponding uncorrected versions.

\subsection{Reconstructed object systematic uncertainties}
\label{sec:experr}
The most important experimental uncertainties originate from the jet mass resolution (JMR) modeling and jet mass scale (JMS) calibration. Uncertainties in \btag efficiency scaling factors and the jet energy scale are found to play a minor role. The remaining uncertainties, including those arising from muon trigger, reconstruction, identification, and isolation rate modeling~\cite{PERF-2015-10}, are negligible.
 
With appreciable reconstructed \Vboson\ and top quark resonance peaks, adjustments of the JMS and JMR central values and uncertainties are possible. Considerations about the validity of transferring these improvements between processes, or along jet \pT, inform the correlation scheme described below.
It was verified that the chosen configuration is conservative in terms of the expected sensitivity to the \Vjets and \Hboson signal strengths.
Jet observables in the simulation are smeared to assess the impact of scale and resolution uncertainties.
A ratio of calorimeter-based to track-based measurements in dijet data and simulation defines the uncertainties in the jet energy (relative 1\%--2\%) and mass scales (relative 2\%--10\%)~\cite{JETM-2018-02}. 
Jet energy scale and mass scale uncertainties are divided into 23 and 6 separate components, respectively, to account for different sources of uncertainty.
The level of JMS agreement between data and simulation, while within the systematic uncertainties, displays a process and jet-\pT\ dependence.
Therefore, JMS uncertainties for \ttbar\ events are separated from those for \Vjets\ and \Hboson\ events within the global likelihood, discussed in Section~\ref{sec:stats}. The dominant component in terms of reconstructed mass scale is further separated so as to act independently on all processes (\ttbar, \Vjets, and \Hboson) and in all analysis jet-\pT\ bins.
Consistent with previous studies of trimmed jets~\cite{STDM-2011-19,ATLAS-CONF-2016-035},
the energy resolution has an absolute 2\% uncertainty, while the mass resolution has a relative 20\% uncertainty.
JMR uncertainties act independently on each process (\ttbar, \Vjets, and \Hboson) and in each analysis jet-\pT\ bin
to account for generator, process, and \pT\ dependence.
The \Vjets\ JMR uncertainty is reduced using independent measurements as described in Section~\ref{sec:VJet}.
 
The impact of uncertainties in \btag rates for $b$-, $c$-, and light-flavor jets is determined separately in various kinematic regions~\cite{FTAG-2018-01, ATLAS-CONF-2018-001, ATLAS-CONF-2018-006}.
Each flavor-category uncertainty is decomposed into independent components.
A specific component for each jet flavor, based on the impact of experimental and theoretical uncertainties,
accounts for an extrapolation of the scaling factors to jets with \pT\ beyond the calibration dataset's kinematic reach~\cite{ATL-PHYS-PUB-2021-003}.
The thresholds are 250~\GeV, 140~\GeV, and 300~\GeV\ for $b$-, $c$- and light-flavor \trkjets, respectively.


\section{Event selection and categorization}

Events are classified into three orthogonal regions: a signal region (SR) used to extract the signal strength, a control region used to study top quark events (\CRtt), and a validation region (VR) used to study the multijet and \Vjets\ background models.
Each region is further configured depending on jet \pT to determine the various signal strengths as described below.

\subsection{Signal and validation regions}
\label{sec:SRandVR}
A uniform requirement for both the VR and SR in all data-taking years of at least one jet with $\pTU>450~\GeV$ and $\mJU>60~\GeV$ removes the kinematic regime biased by the trigger requirements. A second jet with $\pTU>200~\GeV$ is required. At least one of the two leading-\pT\ jets must satisfy the candidate jet criteria.
 
The event categorization first considers the leading jet. If it is a double-tagged candidate jet, the event populates the leading-jet signal region (SRL).
Only if the leading jet is not a double-tagged candidate jet, and the subleading jet is, the event is categorized into the subleading-jet signal region (SRS).
Approximately 40\% of the \Hboson\ events surviving the kinematic cuts pass the $b$-tagging requirement~\cite{ATL-PHYS-PUB-2020-019}.
In simulated multijet events satisfying the SR requirements, roughly 70\% of the candidate jets with mass close to \mH\ contain two $b$-hadrons and less than 5\% of candidate jets do not contain any heavy-flavor hadrons.
The SRS has a sensitivity approximately 50\% lower than that of the SRL, due in part to inferior mass resolution.
 
Both the leading and subleading jets are always considered when creating the validation regions.
The leading-jet validation region (VRL) includes events where the leading jet is an anti-tagged candidate jet, and the subleading jet either has the same distinction or is not a candidate jet.  An analogous definition defines the subleading-jet validation region (VRS). Figure~\ref{fig:eventwisecat} summarizes the event categorization.
 
The signal strength is extracted in an inclusive signal region containing candidate jets with $\pT>250~\GeV$. For the cross-section measurements, the signal region is further configured into a fiducial signal region containing candidate jets with $\pT>450~\GeV$ and four differential signal regions defined by requiring the candidate jet \pT to be in the ranges 250--450~\GeV, 450--650~\GeV, 650--1000~\GeV, or above 1~\TeV.
Only the leading-jet SR is used for the highest-\pT differential SR, $\pT>1~\TeV$.
Only the subleading-jet SR is populated for regions with $\pT<450~\GeV$.
The VRs follow the same definition.
Table~\ref{tab:SRs} summarizes the analysis signal regions.
For measurements within the fiducial and differential regions, the signal yields within \textit{volumes} defined by requirements on the generator's event `truth' record are extracted as described in Section~\ref{sec:results}.
 
Within each region, the discriminating variable, \mJ, is studied in 5~\GeV\ bins.
As a simple analytic function is used to model the multijet background, \mJ\ is studied in a restricted range where this background decreases monotonically.
The combination of the $\mJU>60~\GeV$ selection and the presence of a muon and neutrino from semileptonic $b$-hadron decays reduces the acceptance for \mj\ values near the selection threshold.
Since the \btag efficiency diminishes with decreasing angular distance to another hadronic object~\cite{ATL-PHYS-PUB-2014-014}, the low \mJ\ region in the signal region is further sculpted.
The prevalence of each effect determines the minimum \mJ\ requirement.
The SR mass range in each region is either 70~\GeV\ to 210~\GeV\ or 75~\GeV\ to 210~\GeV, depending on the jet being leading or subleading and the \pT range. The VRs use the same \pT\ bins and \mJ\ ranges.

\begin{figure}[tb]
\centering
\includegraphics[width=0.4\textwidth]{fig_02.pdf}
\caption{Diagram showing the event categorization criteria.
The columns (rows) divide events into categories based on when the leading(subleading)-\pT\ jet is not a candidate jet, when neither of the two leading-\pT\ \trkjets\ are $b$-tagged, when one of the \trkjets\ is $b$-tagged, and when both \trkjets\ are $b$-tagged.
SRL and SRS denote the leading-jet and subleading jet signal regions.
VRL and VRS denote the leading-jet and subleading jet validation regions.
\label{fig:eventwisecat}
}
\end{figure}

\begin{table}[tb]
\begin{center}
\caption{A summary of the candidate jet \pT\ requirements is given for the three analysis SR configurations.
Each SR has an associated \CRtt and VR with the same jet \pT\ requirements.
\label{tab:SRs}}
\begin{tabular}{lcc}
\toprule
\multirow{2}{*}{Region} & \multicolumn{2}{c}{Candidate jet \pT[GeV]} \\
&  SRL     & SRS     \\
\midrule
Inclusive                 &  $>$450  & $>$250  \\
\midrule
Fiducial                  &  $>$450  & $>$450  \\
\midrule
\multirow{3}{*}{Differential}  & 450--650,  & 250--450, \\
& ~~650--1000, & 450--650, \\
& $>$ 1000   & ~650--1000 \\
\bottomrule
\end{tabular}
\end{center}
\end{table}


\subsection{\ttbar control region}
A dedicated \ttbar control region, \CRtt, using muon-trigger events, provides data with a high purity of top quark pair events to determine the \ttbar yield in conditions equivalent to those of the SRs.
The reconstructed final state is a top quark and a top antiquark with one semileptonic decay and one hadronic decay, with a minimum angular separation in the transverse plane.
An isolated muon, with $p_{\mathrm{T}}^{\mu}>52.5~\GeV$, close to a jet \Jb\ containing at least one \trkjet\ defines the former, and a jet \Jt\ with at least three \trkjets\ defines the latter.
Both jets are required to have $\pT>250~\GeV$ and exactly one of the leading two (three) \trkjets\ in \Jb\ (\Jt) must be $b$-tagged.
Considering multiple \trkjets\ within \Jb\ improves the identification efficiency of a $b$-quark reconstructed as a large-$R$ jet by 7.5\%.
 
To ensure the kinematics of top quarks reconstructed in the \CRtt resemble those in the SR, only events with $140<m_{\Jt}<200~\GeV$ are considered, removing those where \Jt\ does not contain all the top quark decay products.
Residual differences between the \CRtt\ \Jt\ and SRS candidate jet \pT\ spectra below 450~\GeV\ originate from the different trigger requirements.
The ratio of the jet \pT\ spectrum in simulated \ttbar\ events in the \CRtt to the jet \pT\ spectrum in the SRS reproduces the difference. It is used to adjust the simulated \ttbar\ event weights and remove data events in the \CRtt with $\pTJt<450~\GeV$.
 
The same \pT boundaries used in the SR are also applied to $p_{\text{T}}^{\Jt}$ to define the inclusive and differential \CRtt\ regions.
The selection criteria, summarized in Table~\ref{tab:CRttSel}, achieve over 95\% purity in \ttbar\ events
for each \CRtt\ \pT\ bin.

\begin{table}[t]
\begin{center}
\caption{A summary of the \CRtt\ selection criteria.
\label{tab:CRttSel}}
\begin{tabular}{lcccc}
\toprule
Jet & $N$ \trkjets & $N$ $b$-tags & Angular selection & Jet mass [GeV] \\
\midrule
\Jb\ & $\ge1$ & 1 & $0.04+10/\pT^{\mu}<\Delta R(\mu,J^b)<1.5$ & -- \\
\Jt\ & $\ge3$ & 1 & $\Delta\phi(J^b,J^t)>\frac{2\pi}{3}$ & 140--200 \\
\bottomrule
\end{tabular}
\end{center}
\end{table}


\FloatBarrier


\section{Higgs boson modeling}

The limited number of event selection criteria pertaining to properties of the recoil system or other activity in the event result in an inclusive analysis in terms of the Higgs boson production modes.
Table~\ref{tab:HiggsModes} shows the relative contributions of the four main production modes as a function of Higgs boson candidate \pT, according to SM predictions and within the Higgs boson window ($105 < \mJ < 140~\GeV$).
In both the SRL and SRS, \ggF\ production contributes the most for $\jetpT>450$~\GeV.
For $\jetpT<450$~\GeV, \ttH\ accounts for around 40\% of the selected Higgs boson events.
A hadronically decaying top quark can satisfy the jet trigger requirements without a high \pTH value, thus resulting in a significant contribution of \ttH\ events with relatively low Higgs boson \pT.
Almost 90\% of \ttH\ events in the Higgs boson window arise from \Hbb decays.
The majority of the remainder are \HWW events and these climb to almost 15\% for larger \mJ values.

\begin{table}[tb]
\caption{The fractional contribution of each production mode to a given analysis region around the Higgs boson mass peak, defined as $105 < \mJ < 140~\GeV$. The fraction is given relative to the total signal yield in the analysis region in question.
\label{tab:HiggsModes}}
\centering
\begin{tabular}{lcccc}
\toprule
\multirow{2}{*}{Process} & \multicolumn{4}{c}{Jet \pT range [GeV]} \\
& 250--450 & 450--650 & 650--1000 & $>1000$ \\
\midrule
\multicolumn{5}{c}{SRL} \\
\midrule
\ggF & -- & 0.56 & 0.50 & 0.39 \\
\VBF & -- & 0.17 & 0.16 & 0.17 \\
\VH  & -- & 0.14 & 0.18 & 0.25 \\
\ttH & -- & 0.13 & 0.16 & 0.19 \\
\midrule
\multicolumn{5}{c}{SRS} \\
\midrule
\ggF & 0.28 & 0.46 & 0.43 & -- \\
\VBF & 0.07 & 0.19 & 0.21 & -- \\
\VH  & 0.26 & 0.24 & 0.26 & -- \\
\ttH & 0.39 & 0.11 & 0.10 & -- \\
\bottomrule
\end{tabular}
\end{table}


The uncertainty on the cross section and acceptance for \ggF-produced events is 20\%. It includes variations of the factorization and renormalization scales, PDF, and parton shower model to account for their respective uncertainties.
Reference~\cite{Lindert:2018iug} demonstrates that the NLO correction is nearly the same in the infinite top-mass approximation and full SM calculation, so no additional systematic uncertainty is assigned.
Uncertainties on the cross section and acceptance for the \VBF, \VH, and \ttH\ processes are taken to be 0.5\%, 5\%, and 13\%, respectively~\cite{Becker:2020rjp}.
Systematic uncertainties in the EW corrections (expressed as $1+\delta_{\text{EW}}$) are taken as $\delta_{\text{EW}}^2$ following the recommendations in Ref.~\cite{deFlorian:2016spz}.


\section{Background process modeling}
\label{sec:background}

Multijet production is the dominant background process.
The \Vjets and top quark resonance peaks flank the Higgs boson signal in low- and high-mass sidebands, respectively, but also leak into the Higgs boson signal window.
Within $105<\mJ<140~\GeV$, \Vjets accounts for about 1\% of the total background, top quarks for about 3\%, and multijets provide the rest.
The expected Higgs boson signal contribution corresponds to 0.2\% of the background for jet $\pT>450~\GeV$ in the mass window and 20\%--80\% of the data statistical precision in the analysis jet-\pT\ bins.
 
Therefore, an accurate and precise determination of the background is paramount and is achieved starting with the determination of the \Vjets and top quark backgrounds.
These backgrounds are determined by maximizing a binned likelihood function defined as the product of Poisson probability terms for each bin of the \mJ\ distributions in the SRL, SRS, and \CRtt.
Unconstrained normalization parameters are common to the signal and control regions within each jet \pT range.
Systematic uncertainties are included as constrained nuisance parameters.
Details are given in Section~\ref{sec:stats}.

\subsection{Top quark production}
The candidate jet in SR top quark events usually contains the remnants of a $b$-quark and the two hadronic decay products of a \Wboson\ boson.
As discussed above, the \CRtt design ensures the same physics processes also populate the control region.
The simulated jet mass distributions in the CR and SR are similar in shape and peak near the top quark mass
because both regions probe a comparable top quark momentum range.
Therefore, any adjustment of the simulated top quark events made to improve their agreement with data in the \CRtt can be directly applied to the SR.
This is achieved by including the \CRtt in the global likelihood described in Section~\ref{sec:stats}.
The inclusive \CRtt has a \ttbar\ purity of 97\% with similar levels found in the fiducial and differential region configurations.
With such high purity, the \ttbar normalization is determined directly from data with better than 10\% precision in most regions.
 
In the SR, $tW$-produced events where a top quark is matched to a candidate jet contribute 2\%-3\% relative to the \ttbar\ yield. For t-channel production, the ratio to \ttbar\ is 1\%--5\%.
Both have a candidate-jet mass distribution similar to that in \ttbar\ production. 
To reduce effects due to limited MC sample size, the mass spectrum's shape for events with one top quark used in the likelihood described in Section~\ref{sec:stats} is obtained by scaling the \ttbar\ MC sample to the number of events predicted by the dedicated $tW$ and t-channel MC samples within each jet \pT\ bin.
The contribution from s-channel production is negligible.
 
For \ttbar\ production, comparisons between nominal and alternative simulated samples provide systematic uncertainty estimates for the parton shower model (\HERWIG[7] replaces \PYTHIA[8]) and the matrix element calculation (\MGNLO replaces \POWHEGBOX[v2]). The comparisons show a 6\%--20\% and 1\%--19\% difference in yield in the various analysis regions, respectively.
Within the nominal sample, variations of internal weights are used to estimate the systematic uncertainties associated with initial- and final-state radiation (1\%--7\%), as well as the renormalization and factorization scales (negligible).
All experimental uncertainties described in Section~\ref{sec:experr} are utilized.
Uncertainties in the \btag efficiency for $b$-jets and the JMS have the largest impact on the \ttbar\ normalization.
A 50\% normalization uncertainty is applied to the estimated number of events with a single top quark, mainly driven by the comparison between the diagram subtraction and diagram removal schemes~\cite{Frixione:2008yi} in $tW$ events.
 
Figure~\ref{fig:topCRMasses} shows the jet mass distribution for each analysis \pT\ bin in the \CRtt\ after the global likelihood fit described in Section~\ref{sec:stats} in the differential configuration. The simulation, shown with a 68\% confidence level (CL) uncertainty band, agrees well with the data.

\begin{figure}[!htb]
\begin{center}
\begin{subfigure}{0.45\textwidth}
\centering
\includegraphics[width=\textwidth]{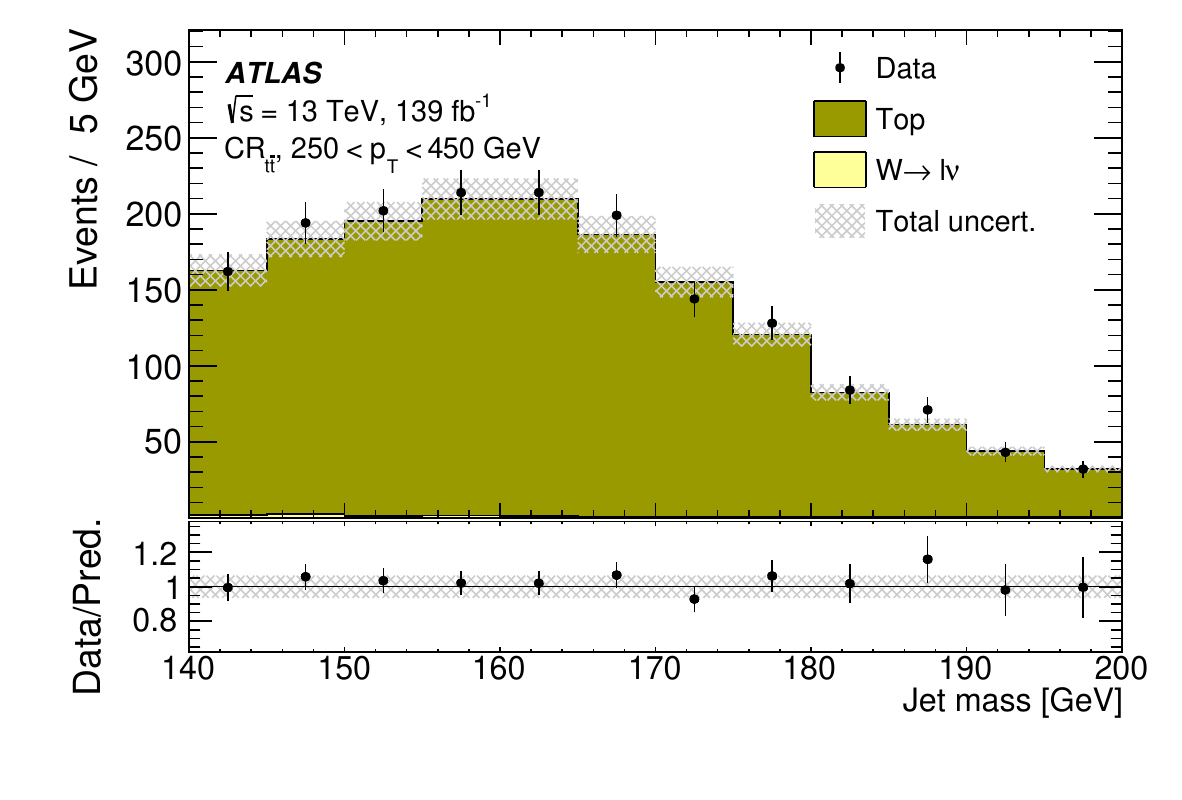}
\end{subfigure}
\begin{subfigure}{0.45\textwidth}
\centering
\includegraphics[width=\textwidth]{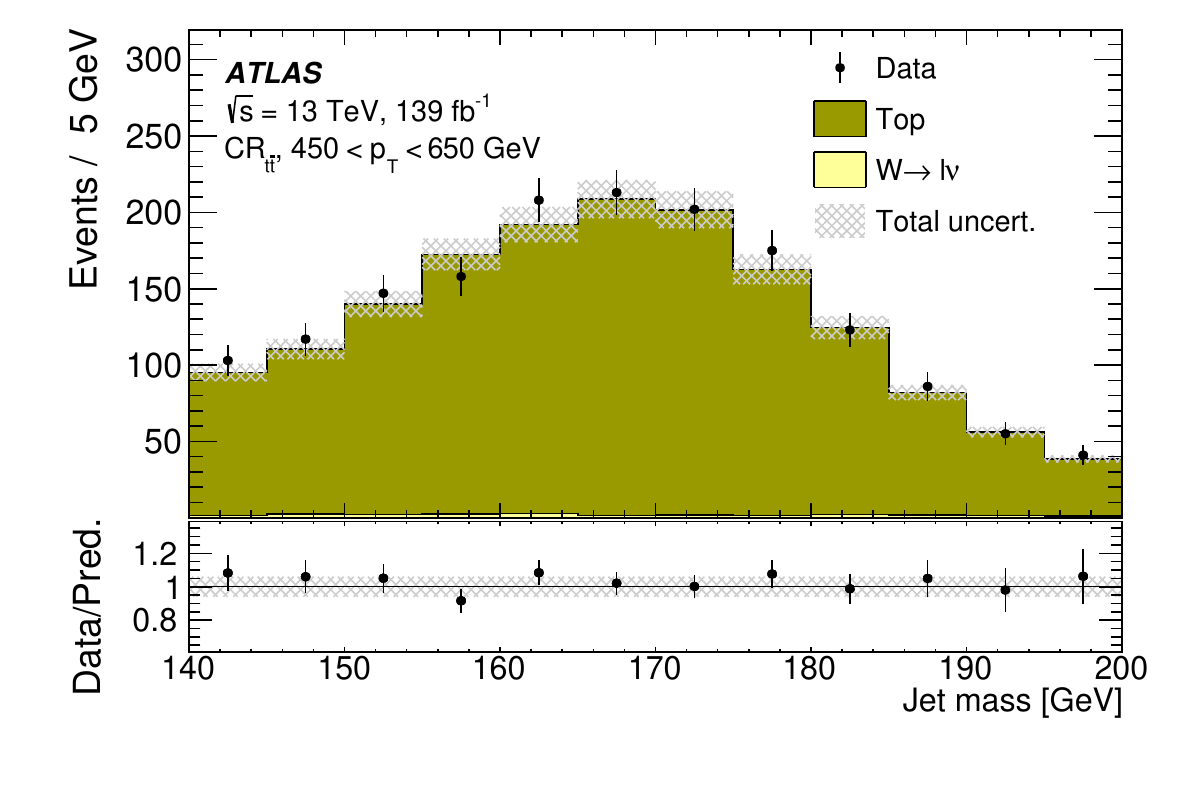}
\end{subfigure}
\begin{subfigure}{0.45\textwidth}
\centering
\includegraphics[width=\textwidth]{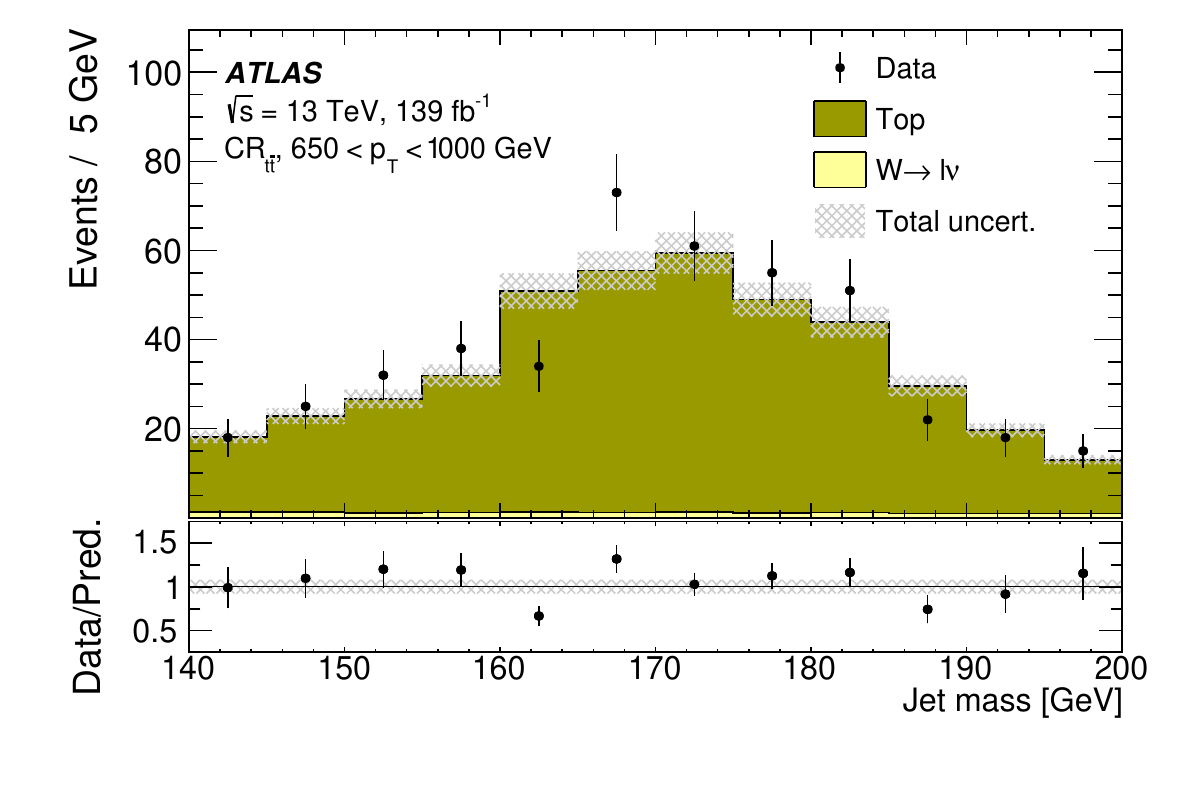}
\end{subfigure}
\begin{subfigure}{0.45\textwidth}
\centering
\includegraphics[width=\textwidth]{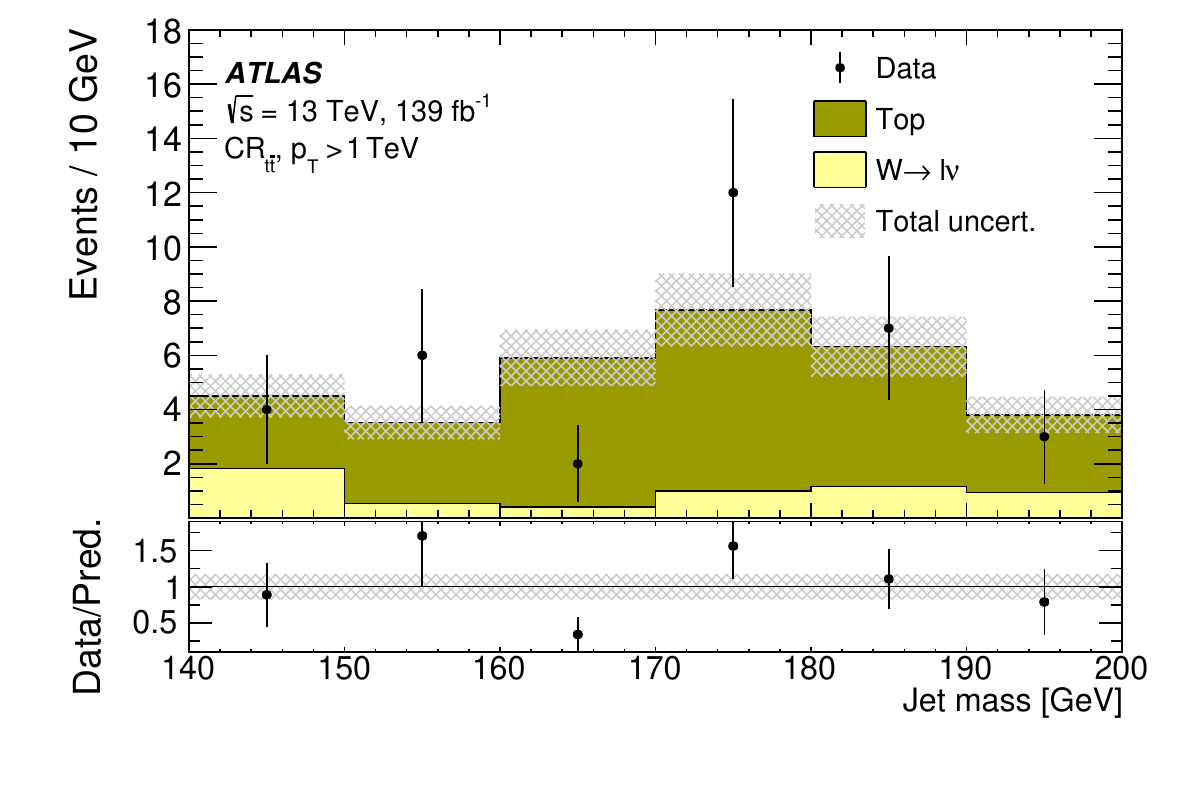}
\end{subfigure}
\end{center}
\caption{The post-fit \CRtt \Jt\ mass distribution in the four \pT regions used in the global likelihood of the differential fit (Section~\ref{sec:stxs}).
The $W(\ell\nu)$ contribution is flat in jet mass and for events with $\pT<1~\TeV$ it is estimated to be 1-3\% of the total.
The $\pT>1~\TeV$ region is shown in 10~\GeV\ jet mass bins.
The ratio of the data to the background prediction is shown in the lower panel.
The shaded areas indicate the 68\% CL for all background processes.
\label{fig:topCRMasses}}
\end{figure}


\FloatBarrier

\subsection{\Vjets\ production}
\label{sec:VJet}
With a decay structure and relative experimental resolution similar to that of the Higgs boson, the vector-boson mass peaks offer a unique opportunity to validate the experimental performance.
Events with \Zboson\ bosons in the signal region outnumber \Hboson\ events by over a factor of 20.
Experimental effects that are challenging to discern in a statistically limited \Hboson\ production measurement will be evident in the \Zboson\ observation. A well-understood \Zboson\ measurement is therefore a precursor to a robust \Hboson\ measurement.
Furthermore, the validation region offers a sample with a topology similar to the SR with which to study \Vjets\ production with a larger event sample but a lower signal-to-background ratio.
 
In the VR, \Wjets\ events outnumber $Z$ events nearly three to one due to the larger cross section and comparable acceptance.
The decay products of the vector boson are reconstructed within the selected candidate jet in over 60\% of events.
In the remainder, the selected candidate jet is created by the recoil hadronic system, resulting in a nonresonant mass distribution, similar in shape to the multijet background, that enhances the high mass tail.
In the SR, the \Zjets\ event contribution is dominant and exceeds that of \Wjets\ events by more than a factor of three, because of the sizable \Zbb branching fraction and the flavor-tagging requirements.
About 90\% of the candidate jets in \Zjets\ events contain the decay products of a vector boson.
Due to the low misidentification rate for \btag, only 40\% of candidate jets in \Wjets\ events are from the boson decay. The prevalence of candidate jets from the recoil system in \Wjets\ events leads to a broad \mJ\ distribution in the SR.
 
The \Zjets normalization is determined directly from the data with the global likelihood described in Section~\ref{sec:stats}.
Therefore, the impact of the considered systematic uncertainties in the modeling is limited to relative changes in acceptance across regions and the \mJ\ distribution shape.
The \Wjets cross section is assigned a 10\% uncertainty in the signal region~\cite{Lindert:2017olm}.
For both processes, the largest effect from seven independent pairs of renormalization and factorization scale variations by factors of 0.5 and 2 corresponds to a 3\%--20\% error in the expected relative acceptance across regions.
An alternative PDF set (\MMHT[nlo]), $\alphas$ variations within the nominal PDF set, and changing the cluster fragmentation model to the Lund string model~\cite{Sjostrand:1984ic} does not lead to a significant change in the acceptance estimate relative to the nominal model.
In the \CRtt, the minor \Wjets\ contribution, referred to as \CRttW, has a total uncertainty of 30\%.
 
All experimental uncertainties described in Section~\ref{sec:experr} are applied. Uncertainties in the JMR and JMS have the largest impact on the \Vjets\ normalization.
Using the multijet model described in Section~\ref{sec:qcd} and the likelihood described in Section~\ref{sec:stats}, the jet mass distribution in the leading-jet validation region is described to the level of agreement between simulation and data shown in Figure~\ref{fig:VRVjFitPtBins}.
 
\begin{figure}[tb]
\begin{center}
\begin{subfigure}{0.40\textwidth}
\centering
\includegraphics[width=\textwidth]{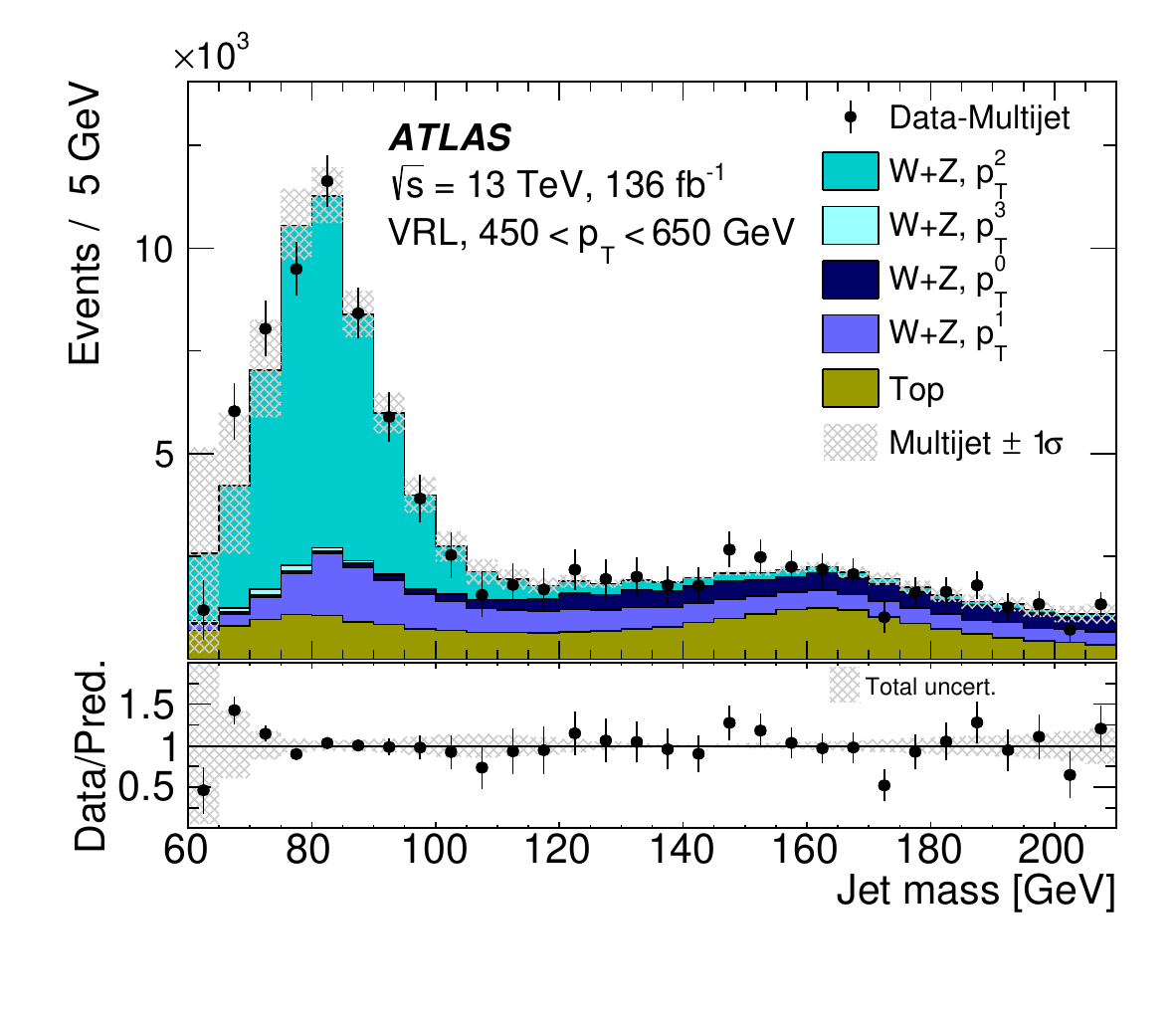}
\caption{$450 < \pT < 650~\GeV$}
\end{subfigure}
\begin{subfigure}{0.40\textwidth}
\centering
\includegraphics[width=\textwidth]{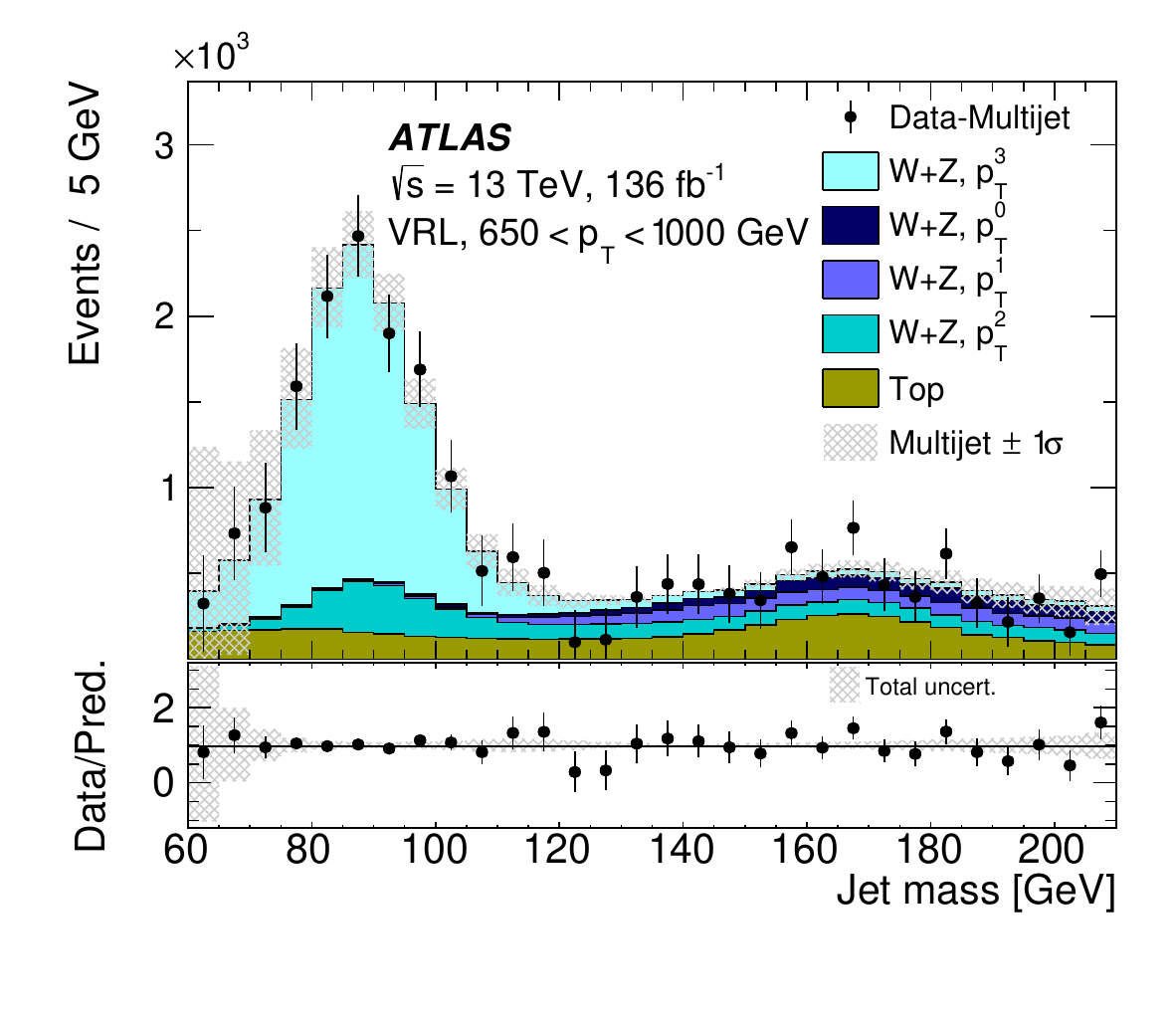}
\caption{$650 < \pT < 1000~\GeV$}
\end{subfigure}
\begin{subfigure}{0.40\textwidth}
\centering
\includegraphics[width=\textwidth]{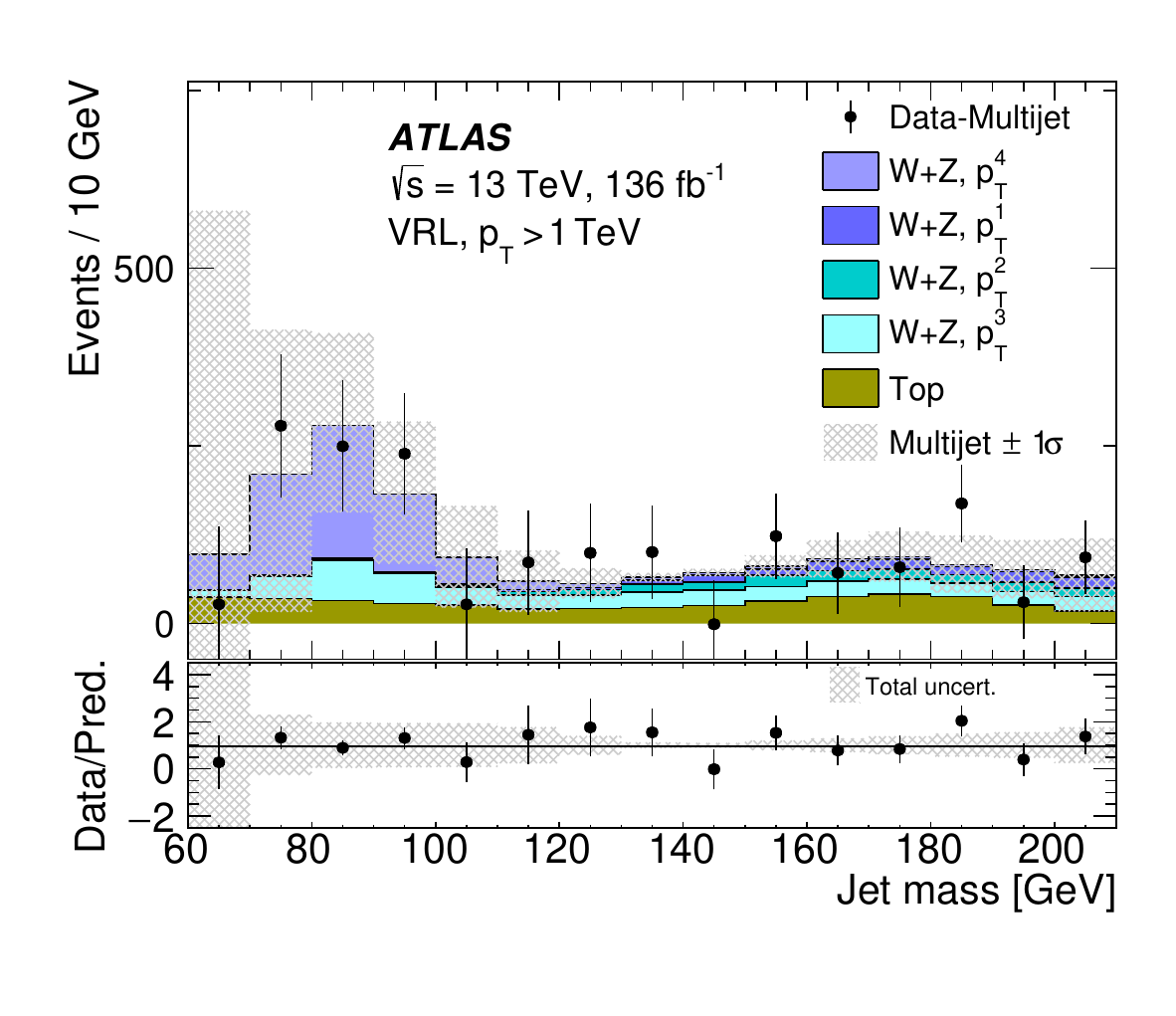}
\caption{$\pT > 1000~\GeV$}
\end{subfigure}
\end{center}
\caption{Post-fit leading-jet invariant mass distributions after the multijet background was subtracted in the validation region for data (points with error bars) and the \Vjets ($W$+$Z$) and \T\ components (histograms) for (a) $450 < \pT < 650~\GeV$, (b) $650 < \pT < 1000~\GeV$, and (c) $\pT > 1~\TeV$ shown in wider 10~\GeV\ jet mass bins.
The \Vjets\ contribution is split into five generator `truth' \pTV\ volumes labeled \pTReg{0}--\pTReg{4} for $\pTV<300~\GeV$, 300--450~\GeV, 450--650~\GeV, 650--1000~\GeV, and ${>}1000~\GeV$, respectively.
The \ttbar normalization and its uncertainty are set to the corresponding values from the \CRtt.
The \mJ\ range has been extended down to 60~GeV for only this fit to show the level of agreement along the rising edge of the \Vjets\ \mJ\ distribution.
The ratio of the data to the background prediction is shown in the lower panel.
The shaded areas indicate the 68\% CL for all background processes.
\label{fig:VRVjFitPtBins}}
\end{figure}

\subsubsection*{Jet mass resolution uncertainty}
The fitted \Zjets\ normalization in the SR had a significant correlation with the reconstructed \mJ\ resolution as the flexibility of the \Zjets\ template and the multijet model could open a local minimum in the likelihood fit minimization procedure.
Tests using subsets of a hybrid validation region, constructed to have a known amount of each process and discussed in the next section, highlighted this feature. In some instances, the best-fit value of the JMR parameter broadened the \Zjets\ peak, corresponding to an increase of the \Zjets\ normalization and a decrease of the multijet contribution compared to the expected values.
 
To stabilize the fit response, the $Z$ and $W$ resonance jet-mass width is measured directly in two data samples as a function of \pT and used to create additional constraints on the \Vjets\ JMR parameter in the global likelihood (see Section~\ref{sec:stats}).
The two data samples are an alternative \ttbar\ CR (\WCRtt) and the VRL. The \WCRtt consists of selected semileptonic \ttbar\ events having a resolved $Wb$ pair from a hadronically decaying top quark, providing a high-purity reconstructed $W$ peak with jet \pT from 200~\GeV\ up to about 600~\GeV. The VRL covers the entire jet-\pT range above 450~\GeV, providing a clear $V$ peak but with considerably more multijet background.
 
\begin{figure}[htb]
\centering
\includegraphics[width=0.5\textwidth]{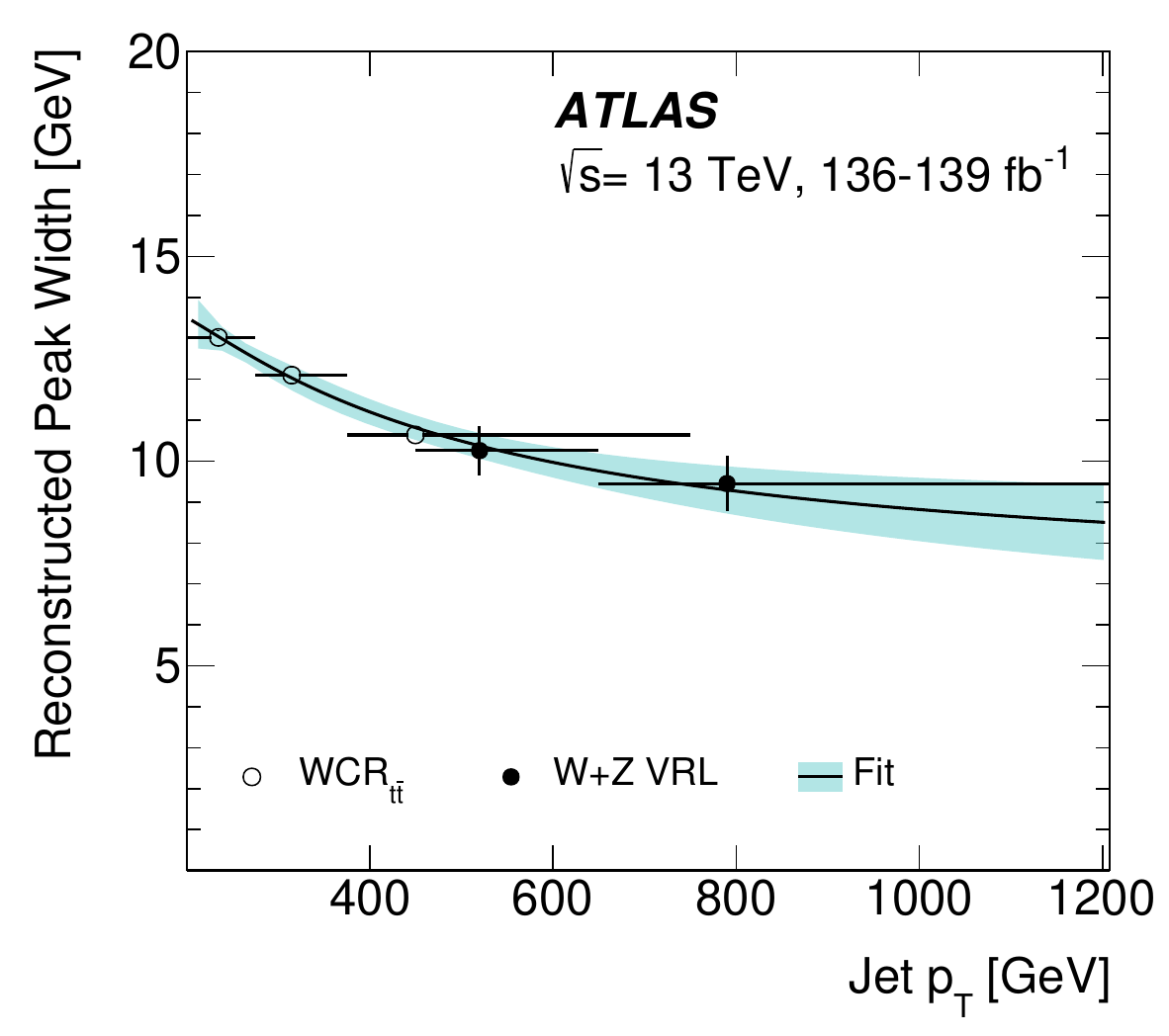}
\caption{A summary of the $Z$ and $W$ resonance peak reconstructed-width measurements as a function of the jet \pT using a resolved $W$ boson in top quark decays in the \WCRtt region and the combined $W$ and $Z$ boson mass distribution in the validation region. The horizontal bars cover the jet \pT range used for the corresponding point and the points are centered at the average jet \pT in the same range.
The continuous black curve is a fit to the measurements with resultant errors shown as a cyan band.
\label{fig:WMassResVsPt}}
\end{figure}
 
The measured jet-mass width of the $W$ and $Z$ resonances shows a smooth evolution from low \pT in the \WCRtt to high \pT in the VRL (see Figure~\ref{fig:WMassResVsPt}).
These results differ from the nominal simulated \mJ\ resolution by less than 2.5\% and have a precision that is around one fifth of the original JMR uncertainty after systematic uncertainties are incorporated to transfer the result to the \Zbb-dominated \Vjets sample in the SR.
When included in the global likelihood, the correlation between the \Zjets normalization and the JMR is reduced compared to when the auxiliary mass measurement is not considered.
For example, in the inclusive signal region it decreases from ${\sim}90$\% to ${\sim}30$\%.
 
\FloatBarrier


\FloatBarrier

\subsection{Multijet production}
The monotonically decreasing jet mass spectrum of the multijet background is modeled with an exponential function of a polynomial:
\begin{equation}
f_{N}\left(x\,\middle|\,\vec{\Theta}\,\right) = \Theta_{0} \displaystyle\exp\left(\sum_{i=1}^{N} \Theta_{i} x^{i}\right),
\label{eq:exppoly}
\end{equation}
where $x = (\mj - 140~\GeV)/{70~\GeV}$ and $\Theta_{i}$ are the parameters of the fit.
Parameter values are determined independently in each region simultaneously with the signal extraction.
The optimal degree of the polynomial, \N in Eq.~(\ref{eq:exppoly}), depends on the mass spectrum's shape and the number of events analyzed.
Values of $N$ which are too large increase correlations between the multijet and resonant process models, decreasing their statistical significance.
Values of $N$ which are too small induce biases in the resonant process yields because the function is too rigid.
These effects are studied in the VRL (VRS), which contains 58 (51) times the amount of SRL (SRS) data.
To determine \N, statistical tests are performed on modified VR subsets, referred to as hybrid VR \textit{slices}, with roughly the same number of events as the corresponding SR.
 
The hybrid VR is the best available proxy for the SR.
The slices are generated from the VR subsets by replacing the VR resonance peaks with the corresponding SM prediction in the SR and correcting the underlying multijet mass spectrum shape to match that of the SR.
The multijet shape difference between the SR and the VR is defined as the ratio of the SR multijet model (\MJSR) to the VR multijet model (\MJVR) (see Figure~\ref{fig:QCDRatioBinned}).
The \MJSR\ is obtained from a global likelihood fit to the SRs and \CRtt, including all experimental and modeling systematic uncertainties.
The Higgs boson signal normalization parameter is blinded and common unconstrained normalization parameters are used within each jet \pT region for the \Vjets\ and \ttbar\ templates separately.
The \MJVR\ is created from the average of the multijet model parameter values obtained from likelihood fits to ten\footnote{Ten subsets balance the statistical precision with the need to use greater values of \N\ to model a larger dataset.} random, orthogonal subsets of the VR including the full set of systematic uncertainties.
The input \ttbar\ normalization and associated uncertainty are set to representative values from the \CRtt.
The VR \Vjets and \T\ estimates (\VVR and \ttVR) are created from the average post-fit contributions from the same ten fits to the VR.
Each slice of the hybrid VR, \VRhi, is built as
\begin{equation*}
\VRhi = (\VRi - \VVR - \ttVR) \times \left(\frac{\MJSR}{\MJVR}\right) + \VSR + \ttSR + \HSR,
\end{equation*}
where \VRi is the jet mass distribution of the data events in the $i^{\text{\tiny th}}$ VR slice and \VSR, \ttSR, and \HSR are the nominal MC predictions for \Vjets, \ttbar, and \Hboson\ production in the SR, respectively.
The ratio of \MJSR\ over \MJVR\ is stable versus $N$.
 
\begin{figure}[!htbp]
\begin{center}
\begin{subfigure}{0.30\textwidth}
\centering
\includegraphics[width=\textwidth]{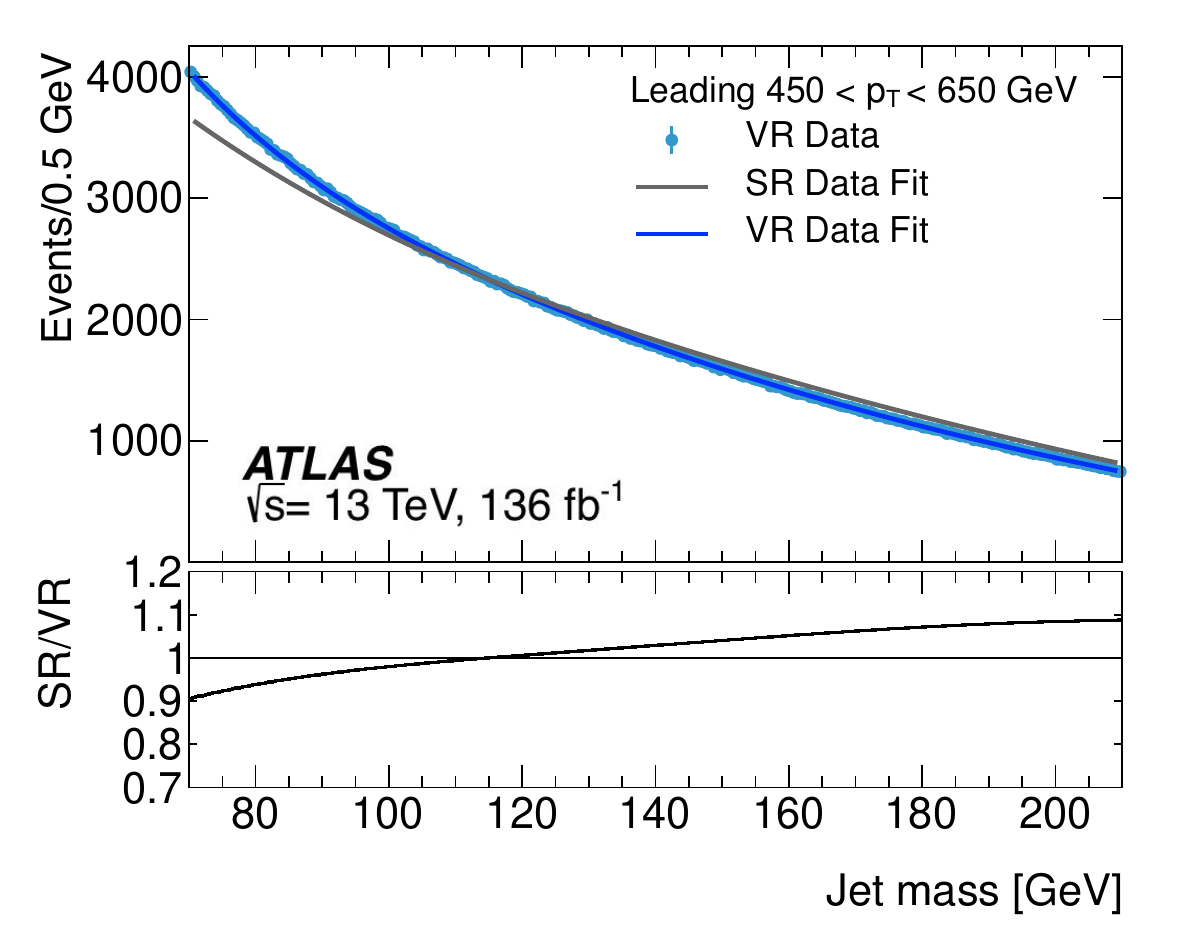}
\caption{SRL: $450 < \pT < 650~\GeV$}
\end{subfigure}
\begin{subfigure}{0.30\textwidth}
\centering
\includegraphics[width=\textwidth]{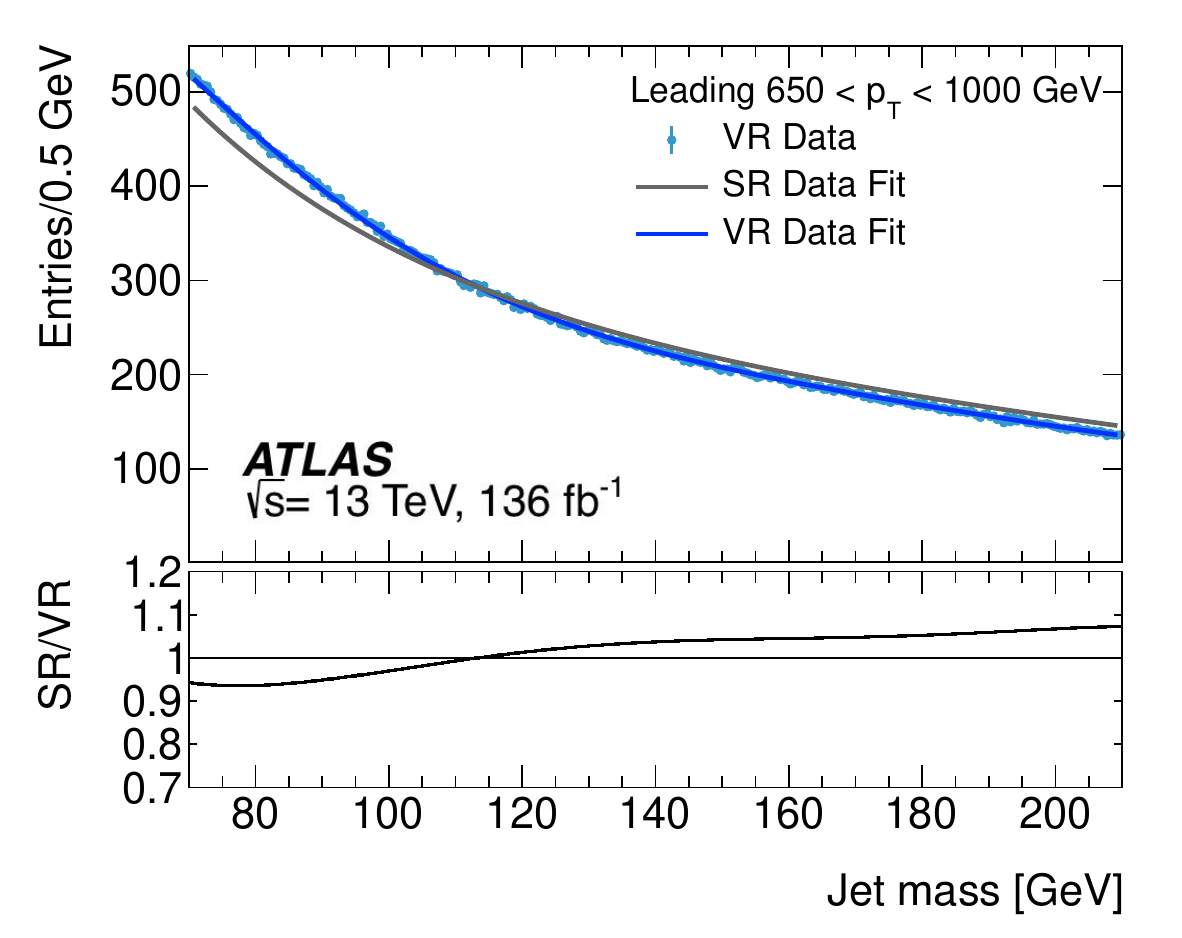}
\caption{SRL: $650 < \pT < 1000~\GeV$}
\end{subfigure}
\begin{subfigure}{0.30\textwidth}
\centering
\includegraphics[width=\linewidth]{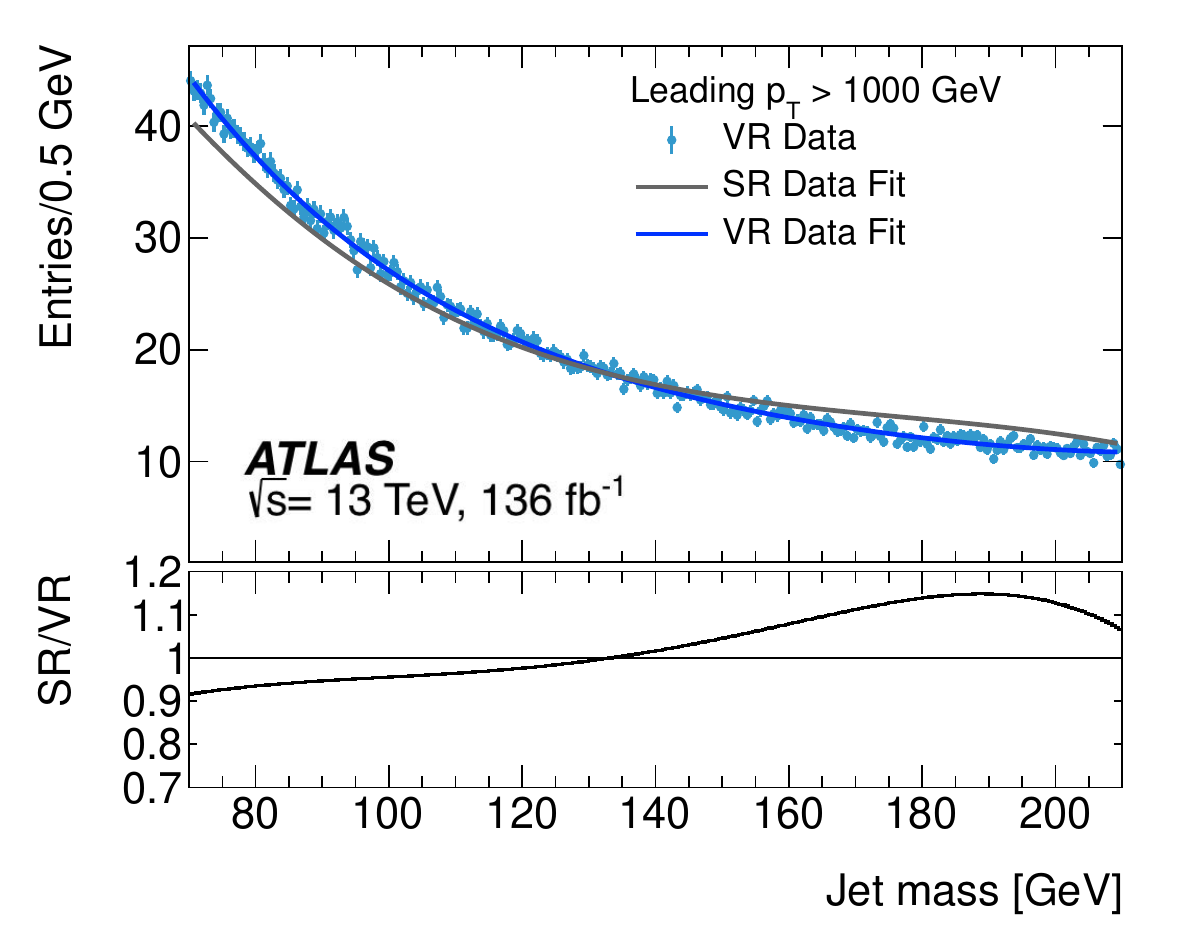}
\caption{SRL: $\pT > 1000~\GeV$}
\end{subfigure}
\begin{subfigure}{0.30\textwidth}
\centering
\includegraphics[width=\textwidth]{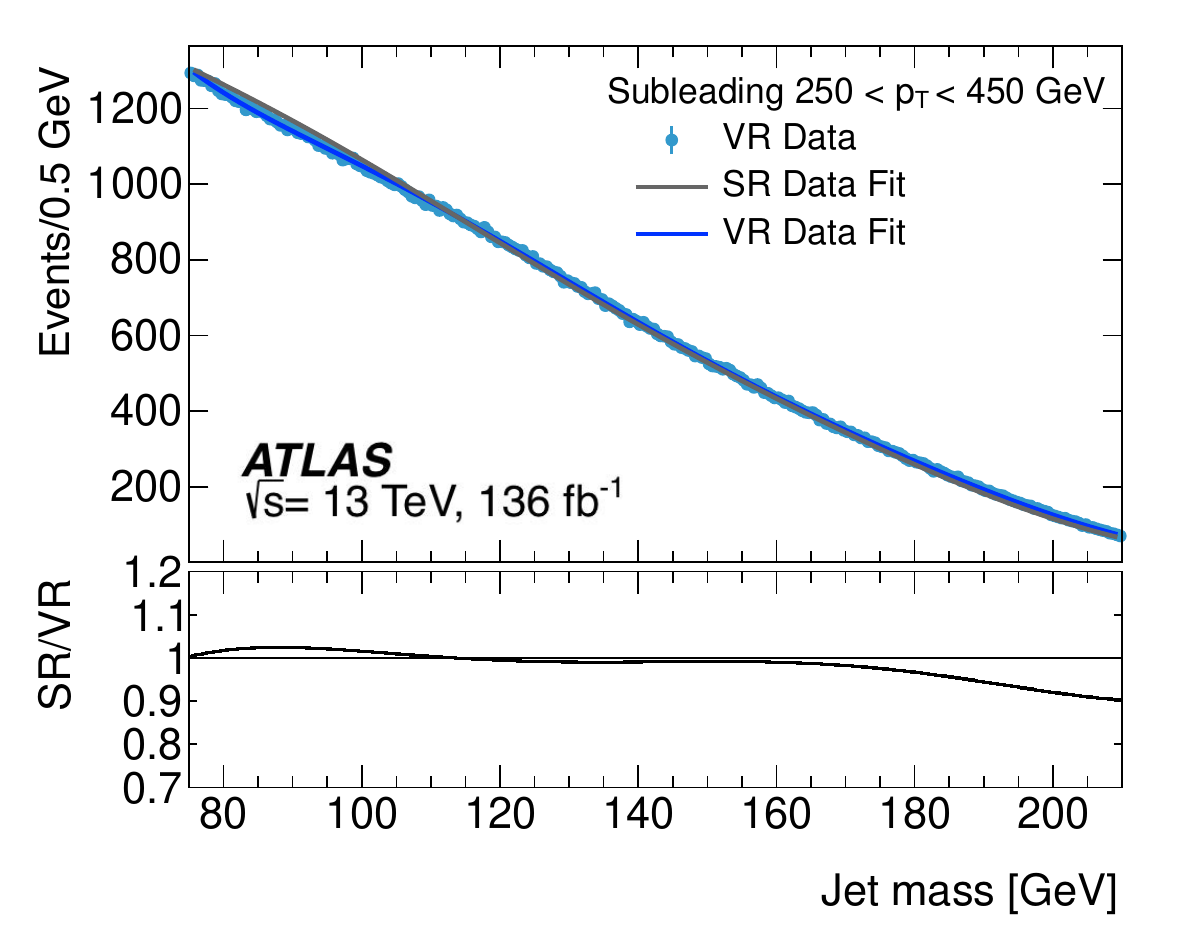}
\caption{SRS: $250 < \pT < 450~\GeV$}
\end{subfigure}
\begin{subfigure}{0.30\textwidth}
\centering
\includegraphics[width=\textwidth]{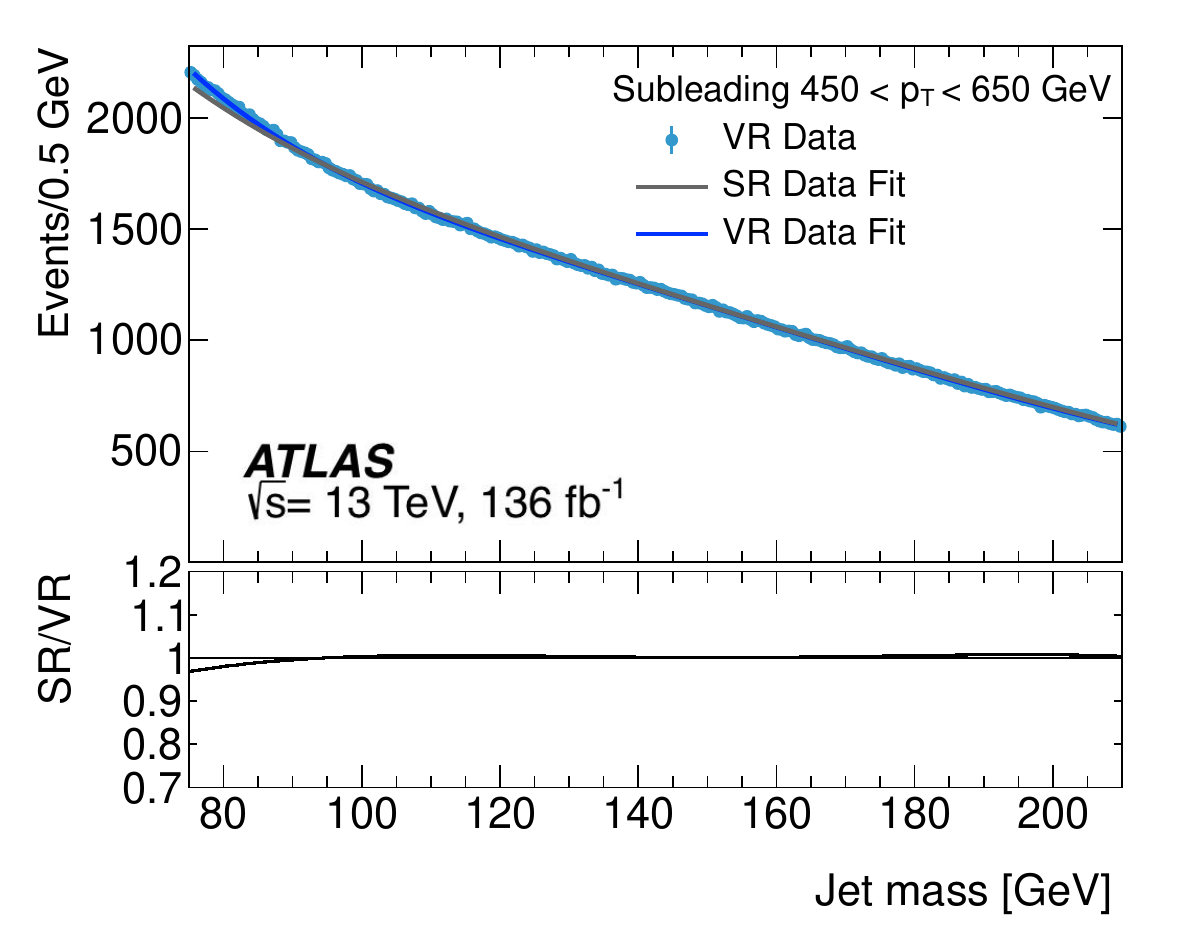}
\caption{SRS: $450 < \pT < 650~\GeV$}
\end{subfigure}
\begin{subfigure}{0.30\textwidth}
\centering
\includegraphics[width=\textwidth]{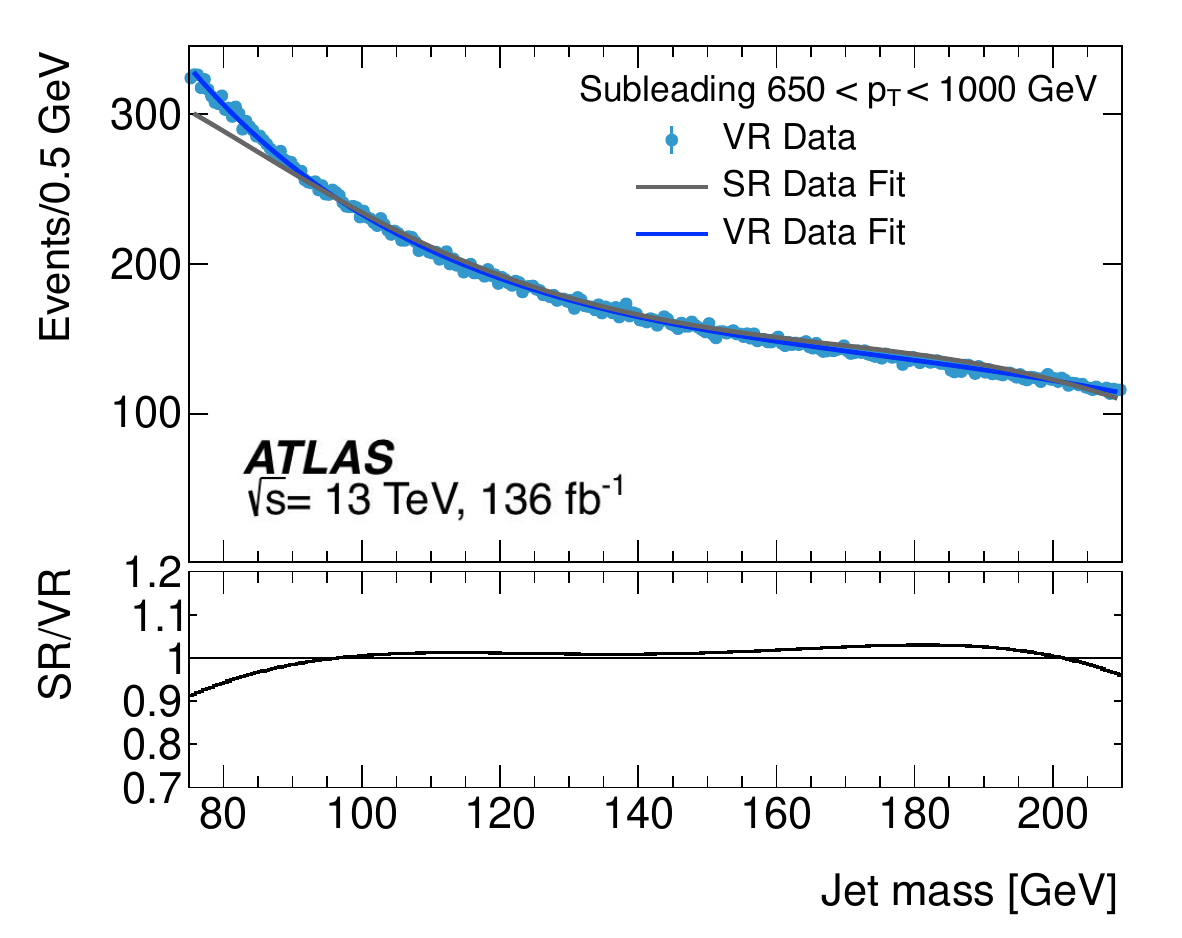}
\caption{SRS: $650 < \pT < 1000~\GeV$}
\end{subfigure}
\end{center}
\caption{Comparison of the multijet jet-mass distribution shapes from the different \pt-binned analysis
regions. The solid lines show the multijet function shape after a fit to the SR data
(gray) and VR data (blue).
The solid points are the data from VR slices with the same number of events as the SR after the SM resonances are subtracted.
The bottom panel shows the ratio of the SR data fit to the VR data fit.
The $250 < \pT < 450~\GeV$ region is only populated in the subleading-jet SR and events
with $\pT>1~\TeV$ are only considered in the leading-jet SR.
\label{fig:QCDRatioBinned}}
\end{figure}
 
The optimal \N for each region is chosen by considering three metrics evaluated with the \VRh collection.
First, a log-likelihood ratio (LLR) test compares the result of an \N-parameter fit (null hypothesis) with the result of an $(N+1)$-parameter fit (alternative hypothesis) in each \VRh slice without any injected SM resonances. By Wilks' theorem, the likelihood ratio follows an asymptotic $\chi^{2}$ distribution with one degree of freedom when the data corresponds to the null hypothesis.
The corresponding distribution of $p$-values is flat.
The smallest \N that yields a uniform distribution of $p$-values is selected.
 
The LLR test ensures a good description of the data over the full mass range,
but resonance measurements are sensitive to local effects.
Two rate tests sensitive to local effects rely on the fit result for a free normalization parameter and its associated error (generalized as \muVR$\pm$\,\sigStatVR) for either the \Zjets\ process 
or the Higgs boson process. 
Both utilize \VRh slices with all SR resonances injected at the SM rates.
The fraction of results \FTS where $|\muVR-1|$ is beyond twice its error \sigStatVR\ gives an estimate of the probability that the multijet model allows a substantial artificial excess or deficit. A $2\sigma$ threshold ensures some results from the full set of VR slices cross the boundary.
The average value of $(\muVR-1)/\sigStatVR$ is denoted by \AvgMu and indicates a bias in determining the signal strength.
The value of \N\ chosen by the LLR test is incremented until \FTS is compatible with 0.05, and
\AvgMu stabilizes
for both \Zjets and \Hboson production.
The chosen values of \N do not change when including systematic uncertainties and when not injecting the \Zjets\ and \Hboson\ processes into the \VRh slices (with appropriate changes to the definition of \FTS and \AvgMu).
A fifth-degree polynomial is used in both inclusive regions, while either a fourth- or fifth-degree polynomial is used in the analysis \pT\ bins.

A nonzero value of \AvgMu for the \N chosen in each region indicates a bias in the background model.
It defines the spurious-signal systematic uncertainty and is in the range 0.01--0.33 for \Hboson and 0.15--0.65 for \Zjets production.
A compatible estimate of the spurious signal systematic uncertainty is found when not injecting the \Zjets\ or \Hboson\ processes into the \VRh slices.
In all cases, the spurious-signal systematic uncertainty has an insignificant impact on the sensitivity.
 
Above 1~\TeV, the \Vjets\ yield extracted from the VR has a sizable statistical uncertainty, necessitating a change in the above procedure (see Figure~\ref{fig:VRVjFitPtBins}).
Comparisons of fit quality and \AvgMu\ for both \Zjets\ and the Higgs boson in the hybrid VR 
for an $N$ of 4, 5, and 6 show no improvement in modeling from the higher-order functions.
The \AvgMu is calculated in two ways to estimate the spurious-signal systematic uncertainty for the chosen value of $N=4$.
The \VVR subtracted from each \VRi\ slice is taken from the average post-fit contributions to five subsets of the VR or the SM prediction.
The spurious-signal systematic uncertainty is found to be 0.5 and 0.3 for \Zjets\ and $H$, respectively, and has an insignificant impact on the sensitivity.
 
Finally, an alternative function for the multijet background modeling, the Laurent series, was tested. It provides a similar level of agreement with the multijet background in the \VRh collection. The differences between the two models are much smaller than the data statistical uncertainty and the expected Higgs boson yield. Therefore, the model choice does not motivate an additional systematic uncertainty.


\label{sec:qcd}


\section{Statistical analysis}
\label{sec:stats}

Signal yields are extracted by minimizing the negative logarithm of a likelihood function $\mathcal{L}(\mu,\vec{\theta})$ with the \textsc{RooStats} framework~\cite{Moneta:2010pm,RooFit}.
The likelihood function is defined as the product of Poisson probability terms, one for each bin of the \mJ\ distribution of the SRL, SRS, and \CRtt.
Bin widths are set to 5~\GeV, necessitating technical advancements within RooStats to fit an analytic function to a wide-binned dataset~\cite{Sciandra}.
Systematic uncertainties enter the likelihood as nuisance parameters, $\vec{\theta}$, constrained with Gaussian or log-normal probability density function priors. The JMR constraints obtained from the \WCRtt\ and VRL regions are included as Gaussian probability density function priors.
Unconstrained, or \textit{free}, parameters control the normalization of the MC templates within each jet \pT region or within a given `truth'-based volume and are common to the SRL, SRS, and \CRtt.
For the multijet model, the function normalization and its polynomial coefficients are free parameters and independent for each jet mass distribution.
Signal yields are expressed as signal strengths, $\mu$, obtained by normalizing the fitted number of signal events to the corresponding SM predictions.
Upper limits on the Higgs boson signal strengths and production cross section are derived using the CL$_\text{s}$ method~\cite{Junk:1999kv,Read:2002hq}. The expected limits assume no Higgs boson contribution.
 
Table~\ref{tab:systs} summarizes the systematic uncertainties considered in the likelihood fit. In addition, uncertainties due to the limited number of events in the simulated samples used for the background predictions are parameterized using the Beeston--Barlow `lite' technique~\autocite{BeestonBarlow}.
Systematic variations yielding large statistical fluctuations are smoothed using custom algorithms which also remove variations resulting in effects below 2\% within a given region.

\begin{table}[tb]
\begin{center}
\caption{A summary of the systematic uncertainties included within the profile likelihood for the \Hboson\ and \Zboson\ signal strength extraction.
For a given uncertainty, the second column lists each process which has independent nuisance parameters within the likelihood.
The third column describes how the systematic uncertainty is correlated across regions: ``all'' indicates a fully correlated parameter, ``\pT bins'' indicates a decorrelation between the analysis \pT\ bins, and ``LS'' means it is decorrelated between the SRL and SRS. For the inclusive and fiducial analysis configurations, ``bins'' does not apply, and should be understood to mean the same as ``all''.
The fourth column describes the change induced by the parameter. ``S'' means the \mJ\ shape will change while ``N'' denotes parameters which change the normalization and can result in a migration of events between regions.
$(\ast)$ Two minor components separately apply to \ttbar\ and \Vjets\ events.
$(\bullet)$ For cross section limits and measurements, this uncertainty covers relative acceptance across regions instead of the absolute cross section uncertainty.
$(\circ)$ The spurious-signal uncertainty is only applied to \Zjets\ when the procedure to extract signal strengths in `truth'-based volumes is tested using \Zjets\ events in the SR.
\label{tab:systs}}
\begin{tabular}{lccc}
\toprule
Description & Processes & Category & Effect \\
\midrule
\multicolumn{4}{c}{Reconstructed object systematic uncertainties} \\
\midrule
JMR               & \ttbar, \Vjets, \Hboson   & \pT bins & N+S  \\
JMS (dominant)    & \ttbar, \Vjets, \Hboson   & \pT bins & N+S  \\
JMS (rest)        & \ttbar, \Vjets$+$\Hboson  & all  & N+S      \\
Jet energy scale  & all$^{(\ast)}$ & all  & N+S \\
Jet energy resolution  & all                       & all  &  N+S \\
$b$-tag efficiency for $b$-jets & all & all & N+S \\
$b$-tag efficiency for $c$-jets & all & all & N+S \\
$b$-tag efficiency for light-flavor jets & all & all & N+S \\
\midrule
\multicolumn{4}{c}{Process modeling systematic uncertainties} \\
\midrule
Renormalization and factorization scale & \Vjets & all & N+S \\
Cross section & \Wjets & all & N   \\
Cross section and acceptance & $W(\ell\nu)$ & all & N \\
Parton shower model               & \ttbar & all & N+S \\
Matrix element calculation        & \ttbar & all & N+S \\
Initial- and final-state radiation    & \ttbar & all & N+S \\
Cross section and acceptance      & $t$ & all & N \\
Cross section and acceptance$^{(\bullet)}$ & \Hboson  & all & N \\
NLO EW corrections & \VBF$+$\VH$+$\ttH & all & N \\
\multirow{2}{*}{Spurious signal}  & \Hboson          & \pTH bins $\times$ LS & N \\
& \Zjets$^{(\circ)}$ & \pTZ bins $\times$ LS & N \\
\bottomrule
\end{tabular}
\end{center}
\end{table}


\FloatBarrier


\section{Results}

The three analysis configurations designed to probe Higgs boson production with considerable transverse momentum are summarized in Table~\ref{tab:SRs}. The inclusive region provides a measure of the \Hboson\ signal strength, the fiducial region allows a measurement of the fiducial cross section, and the differential regions are used to measure the cross section in four \pT bins.
All \Hboson\ production modes are considered for the signal strength extraction.
Signal strengths within the fiducial region consider events within a fiducial \textit{volume} defined by requirements on the generator's event `truth' record, the Higgs boson transverse momentum (\pTH) and rapidity (\yH).
Using the same `truth' information within the differential regions, \textit{volumes} like the bins used in the simplified template cross-section (STXS) framework and simply referred to as STXS volumes, are considered;
the \pTH requirements match those for \ggF\ production in the STXS scheme~\cite{deFlorian:2016spz,Badger:2016bpw,Amoroso:2020lgh}, but the \yH\ requirement is more stringent and all production modes are included.
The analysis jet-\pT\ bins align well with the \pTH-defined volumes.
The yield of signal events outside the targeted volume(s) are constrained to their SM prediction within the theoretical and experimental uncertainties.
The cross sections are obtained from the fitted signal yields divided by the integrated luminosity, corrected by the product of the estimated selection efficiency and fiducial or STXS volume acceptance.
Using the same \pT boundaries, cross-section measurements of \Vjets\ production in the VRL and of \Zjets\ production in the SR validate the method.
For these tests, the fiducial and STXS volumes are defined by the generator `truth' vector-boson transverse momentum (\pTV).
Table~\ref{tab:volumes} summarizes the fiducial and STXS volumes and the SM predicted cross sections are in the HEPData repository~\cite{hepdata}.

\begin{table}[tb]
\begin{center}
\caption{A summary of the fiducial and STXS volumes used to determine which signal events are considered for the signal strength measurement in the fiducial and differential regions, respectively.
Signal cross sections outside of these volumes are constrained to their SM prediction.
\label{tab:volumes}}
\begin{tabular}{lcc}
\toprule
Volume        & \pTH[GeV] & $\absyH$ \\
\midrule
Fiducial &  $>$450    &  $<2$ \\
\midrule
\multirow{4}{*}{STXS}          &      300--450,  & \mrFo{$<2$} \\
&      450--650,  &  \\
&      ~~650--1000, &  \\
&      $>$1000    &  \\
\bottomrule
\end{tabular}
\end{center}
\end{table}


\FloatBarrier
 
\subsection{Inclusive region}

The inclusive region yields a Higgs boson signal strength of $\muH{}=0.8\pm 3.2$ for the combination of SRL, SRS and \CRtt. The results are summarized in Table~\ref{tab:inclFitData} and the yields in Table~\ref{tab:yieldsIncl}.
The post-fit SRL and SRS jet mass distributions are shown in Figure~\ref{fig:SRLInclPostFit}.
The Higgs boson signal sensitivity is limited by the size of the data sample. The leading sources of systematic uncertainty are the jet mass resolution and mass scale.
The \mutt{} value is compatible with \ttbar event measurements in a similar kinematic phase-space~\cite{TOPQ-2018-15}.
With nearly 99\% purity in top-jet events, the \CRtt\ data reduces the top quark jet JMR uncertainty to a relative 7\% (from 20\%).
For \Vjets (\T) events, the JMS uncertainty's nuisance parameter is pulled by $-80\%$ (20\%) of its original width, and its width is reduced by 50\% (50\%).
The \Vjets resonance peak position moves by about 2~\GeV\ from the MC prediction.
No other nuisance parameters are modified significantly.

\begin{table}[tb]
\caption{Expected and observed values of the signal strengths for the $H$, $Z$ and \ttbar components in the inclusive fit.
Absolute cross section uncertainties are not included in the reported \ensuremath{\mu_{Z}}\xspace and \ensuremath{\mu_{\ttbar}}\xspace values.
\label{tab:inclFitData}}
\begin{center}
\begin{tabular}{l
S[table-format= 1.1(2)] 
S[table-format= 1.2(2)] 
S[table-format= 1.2(2)] 
}
\toprule
Result & {\ensuremath{\mu_{H}}\xspace} & {\ensuremath{\mu_{Z}}\xspace} & {\ensuremath{\mu_{\ttbar}}\xspace} \\
\midrule
Expected & 1.0 \pm 3.2  & 1.0  \pm 0.17 & 1.0  \pm 0.07 \\
Observed & 0.8 \pm 3.2  & 1.29 \pm 0.22 & 0.80 \pm 0.06 \\
\bottomrule
\end{tabular}
\end{center}
\end{table}


\begin{table}[tb]
\begin{center}
\caption{Event yields and associated uncertainties after the global likelihood fit in the inclusive region.
The total background yield can differ from the sum of the individual components due to rounding.
\label{tab:yieldsIncl}}
\centerline{
\begin{tabular}{l
S[table-format=  6.0(4)]
S[table-format=  6.0(4)]
S[table-format=  4.0(2)]
}
\toprule
Process & {SRL} & {SRS} & {\CRtt}  \\
\midrule
Multijet & 590700 \pm 4200 & 529300 \pm 3500 & {--}   \\
\Zjets & 16100 \pm 2800 & 12000 \pm 2100 & {--}   \\
\Wjets & 3050 \pm 720 & 2510 \pm 500 & {--}   \\
Top & 16300 \pm 1900 & 15900 \pm 2000 & 3737 \pm 68  \\
$W(\ell\nu)$ & {--}  & {--}  & 53 \pm 16  \\
\Hboson & 400 \pm 1500 & 300 \pm 1300 & {--}   \\
\midrule
Total & 626530 \pm 820 & 560090 \pm 770 & 3790 \pm 66  \\
\midrule
Data & 626532 & 560083 & 3791  \\
\bottomrule
\end{tabular}
} 
\end{center}
\end{table}


\def\SRPostFitCapMid{The middle panels show the distributions after subtraction of the multijet distribution. The shaded areas indicate the 68\% CL for the multijet background from the fitted parameters and normalizations of the exponentiated polynomials. }
\def\SRPostFitCapLower{The lower panels show the distributions after subtraction of all the fitted background processes: multijet, \Vjets, and \T. The shaded areas indicate the 68\% CL for all background processes. }
 
\begin{figure}[htb]
\begin{center}
\begin{subfigure}{0.30\textwidth}
\centering
\includegraphics[width=\textwidth]{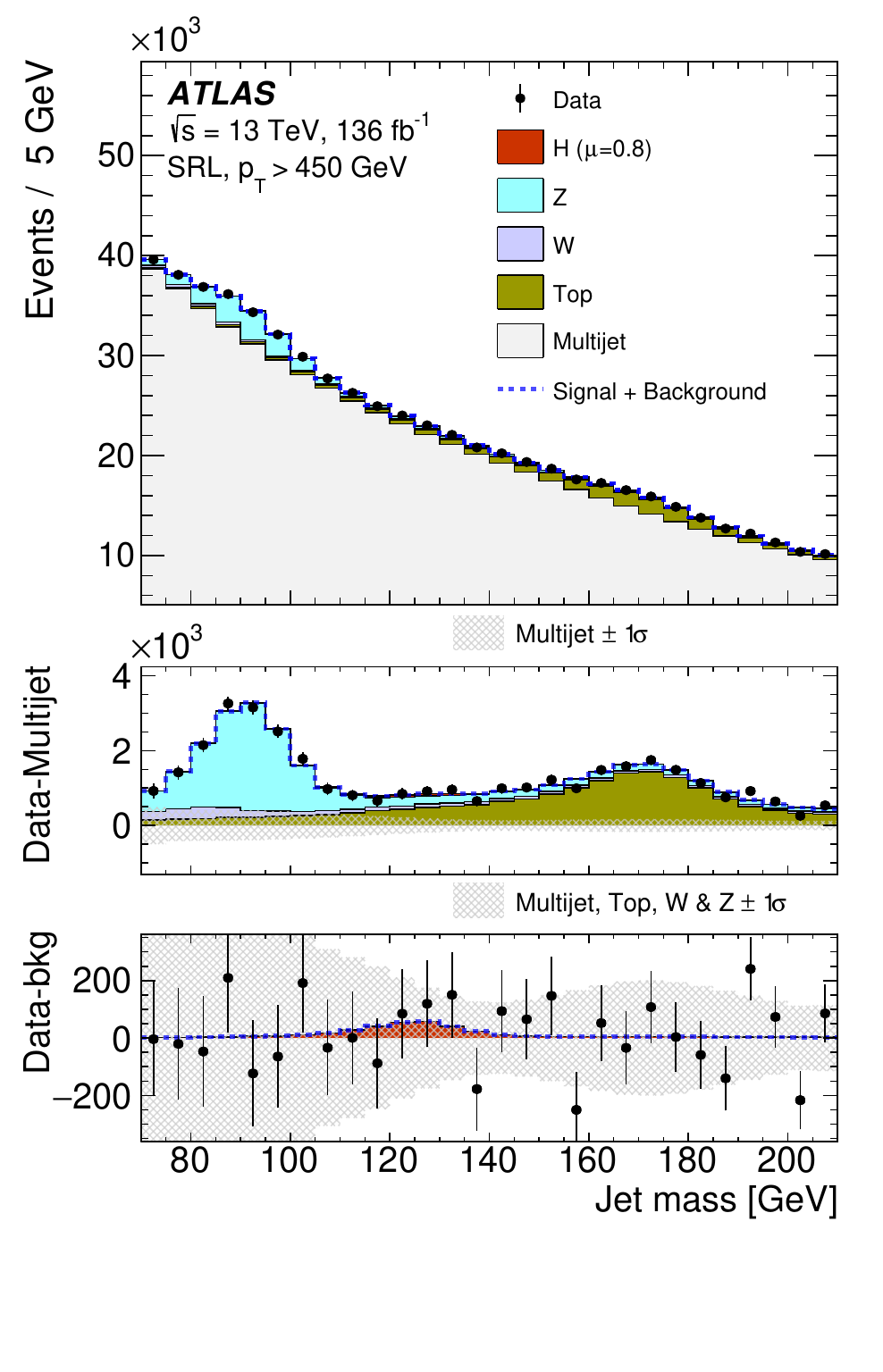}
\end{subfigure}
\begin{subfigure}{0.30\textwidth}
\centering
\includegraphics[width=\textwidth]{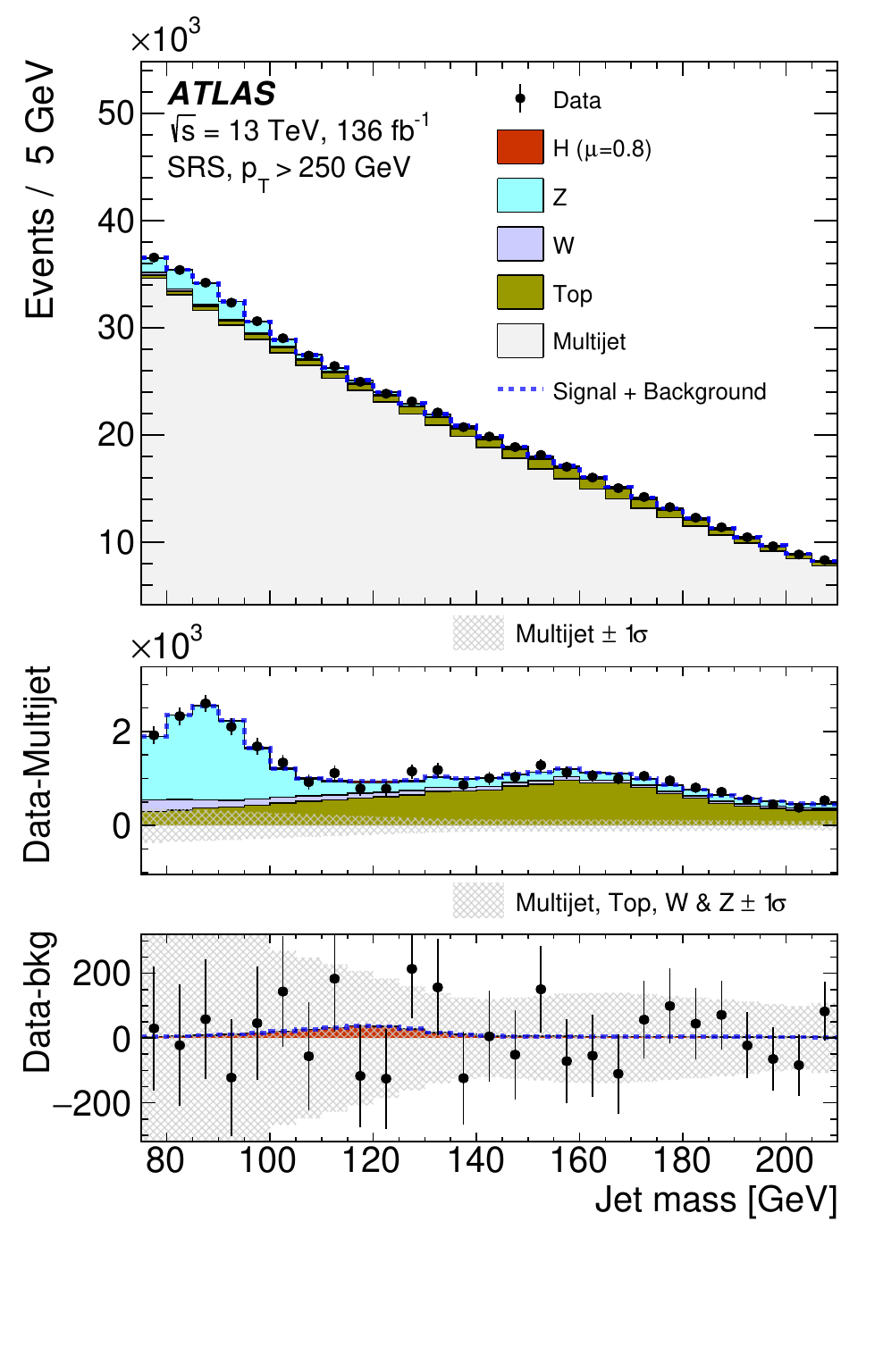}
\end{subfigure}
\end{center}
\caption{Post-fit jet mass distributions for the various components in the inclusive SRL (left) and SRS (right) regions.
\SRPostFitCapMid
\SRPostFitCapLower
\label{fig:SRLInclPostFit}}
\end{figure}


\FloatBarrier
 
\subsection{Fiducial region}

In the fiducial region, the Higgs boson yield and cross section are determined within the fiducial volume defined by the Higgs boson transverse momentum $\pTH>450~\GeV$ and rapidity $|\yH|<2.0$. Compared to the inclusive measurement discussed above, the fiducial region does not include the SRS region below 450~\GeV.
The acceptance times efficiency values for the different SM Higgs boson production processes are given in Table~\ref{tab:fiducialAccept}.

\begin{table}[tb]
\begin{center}
\caption{Signal acceptance times efficiency within the fiducial volume used in the fiducial region.
\label{tab:fiducialAccept}}
\begin{tabular}{lcc}
\toprule
\multirow{2}{*}{Process} & $\pTH>450~\GeV$ \\
& $\absyH<2$ \\
\midrule
All   & 0.24 \\
\midrule
\ggF & 0.26 \\
\VBF & 0.22 \\
\VH  & 0.27 \\
\ttH & 0.20 \\
\bottomrule
\end{tabular}
\end{center}
\end{table}


Two Higgs boson mass templates are used in each SR. The first describes the jet mass distribution of events within the fiducial volume; the second includes events outside the fiducial volume, i.e.\ those with \pTH below 450~\GeV.
The first component accounts for more than 80\% of the Higgs boson signal selected and has a free normalization parameter common to the SRL and SRS. The second component is constrained to the SM value within the theoretical and experimental uncertainties and also tends to have a broader mass spectrum shifted to higher values.
 
This procedure is first tested with \Wqq and \Zqq in the VR and \Zbb in the SR.
For these tests, the $V$ and $Z$ mass templates are structured similarly to those of the Higgs boson described above.
The Higgs boson yields are kept fixed to the SM expectations in the fit and the result is insensitive to this choice.
The fitted signal strength for \Vjets with $\pTV > 450~\GeV$ is $1.01 \pm 0.09$. In the SR, the signal strength for \Zboson events with $\pTZ > 450~\GeV$ is $1.35 \pm 0.25$. These results are in agreement with the SM.
 
When extracting the Higgs boson signal strength, the likelihood fit result for $\pTH>450~\GeV$ provides a signal strength of $\ensuremath{\mu_{H}}\xspace = -0.1 \pm 3.5$.
The corresponding observed (expected) 95\% CL upper limit on the Higgs boson production cross section is
\begin{equation*}
\sigma_{H}(\pTH>450~\GeV) < 115~(128)~\mathrm{fb} \\
\end{equation*}
and the SM prediction is 18.4~fb.

The results are summarized in Table~\ref{tab:fiducialData}.
The \mutt{} value is compatible with \ttbar event measurements in a similar kinematic phase-space~\cite{TOPQ-2018-15}.
The Higgs boson signal strength sensitivity is limited by the data sample size, and the impact from the main sources of uncertainty is given in Table~\ref{tab:fiducialSysts}.
The jet uncertainties give the largest contribution of systematic uncertainty, driven by JMS effects.
Half of the JMS contribution comes from the JMS uncertainty for both the \Vjets\ and \ttbar\ events, while the other half corresponds to the uncertainty in the JMS of reconstructed Higgs bosons.
With nearly 99\% purity in top-jet events, the \CRtt\ data reduces the top quark jet JMR uncertainty to a relative 5\%, one quarter of the original size.
In \Vjets\ events, the dominant JMS uncertainty's nuisance parameter is pulled by $-80\%$ of its original width, and its width is reduced by 50\%. This moves the \Vjets resonance peak position about 2~\GeV.
For \T\ events, the corresponding parameter is also constrained to 50\% of its original value.
No other nuisance parameters are modified significantly.

\begin{table}[tb]
\caption{Expected and observed values of the signal strengths for the \Hboson, \Zboson, and \ttbar\ components in the fiducial fits. The value for \ensuremath{\mu_{H}}\xspace refers to the fiducial volume $|\yH|<2.0$ and $\pTH>450~\GeV$, while those for \ensuremath{\mu_{Z}}\xspace and \ensuremath{\mu_{\ttbar}}\xspace pertain to regions of jet \pT above 450~GeV.
Absolute cross section uncertainties are not included in the reported \ensuremath{\mu_{Z}}\xspace and \ensuremath{\mu_{\ttbar}}\xspace values.
\label{tab:fiducialData}}
\begin{center}
\begin{tabular}{l
S[table-format=-1.1(2)] 
S[table-format= 1.2(2)] 
S[table-format= 1.2(2)] 
}
\toprule
Result & {\ensuremath{\mu_{H}}\xspace} & {\ensuremath{\mu_{Z}}\xspace} & {\ensuremath{\mu_{\ttbar}}\xspace} \\
\midrule
Expected &  1.0 \pm 3.4  & 1.0  \pm 0.18 & 1.0  \pm 0.08 \\
Observed & -0.1 \pm 3.5  & 1.30 \pm 0.22 & 0.75 \pm 0.06 \\
\bottomrule
\end{tabular}
\end{center}
\end{table}


\begin{table}[tb]
\caption{Contributions to the systematic uncertainties for the measurement of the fiducial volume signal-strength, defined as the signal yield relative to the SM prediction. The total uncertainty is also given for comparison.
\label{tab:fiducialSysts}}
\begin{center}
\begin{tabular}{l S[table-format=2.1] }
\toprule
Uncertainty Contribution & {$\pTH > 450\, \GeV$} \\
\midrule
Total                       & 3.5 \\
\midrule
Statistical                 & 2.6 \\
Systematic                  & 2.3 \\
\midrule
Jet systematic uncertainties             & 2.2 \\  
Modeling and theory systs.  & 0.8 \\
Flavor-tagging systs.       & 0.2 \\
\bottomrule
\end{tabular}
\end{center}
\end{table}



\FloatBarrier
 
\subsection{Differential regions}
\label{sec:stxs}

Possible deviations from the SM predictions could manifest with an amplitude increasing with \pTH.
The differential regions aim to measure the Higgs boson transverse momentum spectrum in four \pTH volumes 300--450~\GeV, 450--650~\GeV, 650--1000~\GeV, and greater than 1~\TeV\ using four signal regions requiring the candidate jet \pT to be in the range 250--450~\GeV, 450--650~\GeV, 650--1000~\GeV, or above 1~\TeV\ (see Table~\ref{tab:SRs}).
Extending the procedure adopted for the fiducial region measurement, a Higgs boson mass template for each \pTH volume is used within each jet \pT region in the global likelihood.
The $\pTH>1~\TeV$ volume probes a new domain of highly boosted $Z$ and Higgs boson reconstruction. The expected sensitivity in the SRS with jet \pT above 1~\TeV\ is marginal because the muon-in-jet correction and $b$-tagging turn-on effects are more significant than in the SRL.
Therefore, only the SRL and \CRtt\ regions are included for measurements above 1~\TeV.
Figure~\ref{fig:STXSNumbers} presents the expected signal yield in each candidate jet \pT\ region for each STXS volume and the corresponding fraction of signal events.
The acceptance times efficiency values for the different Higgs boson production processes are given in Table~\ref{tab:STXSAccept}.
 
\begin{figure}[tb]
\begin{center}
\begin{subfigure}{0.45\textwidth}
\centering
\includegraphics[width=\textwidth]{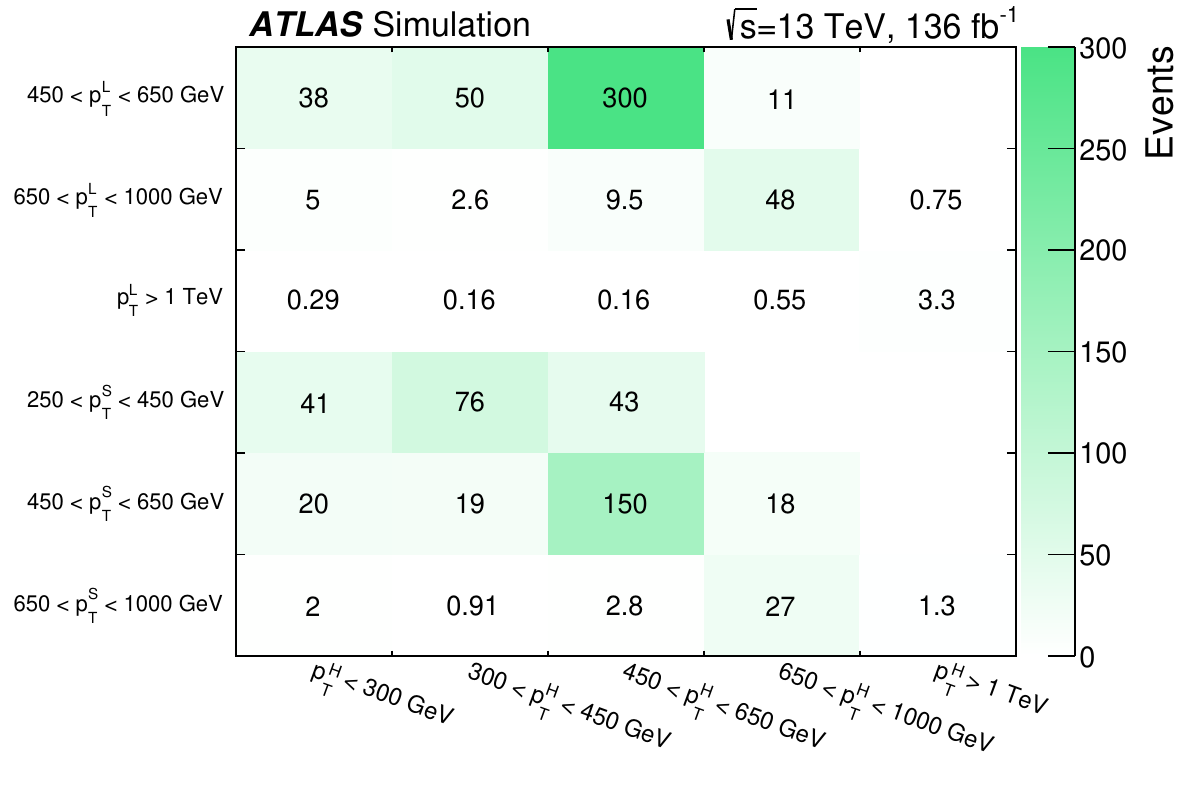}
\end{subfigure}
\begin{subfigure}{0.45\textwidth}
\centering
\includegraphics[width=\textwidth]{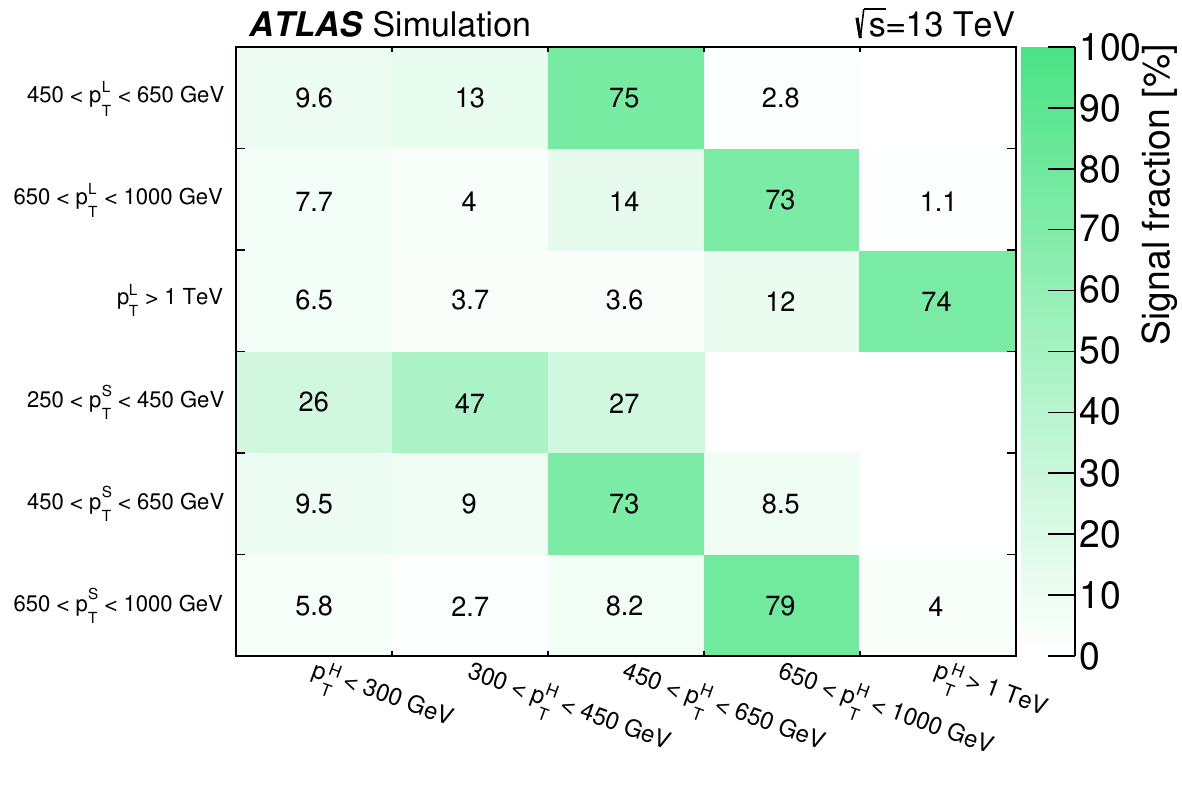}
\end{subfigure}
\end{center}
\caption{
For each of the \pTH\ differential volumes ($x$-axis),
the expected signal event yield for all Higgs boson events (left) and the fraction of signal in percent (right) in each reconstructed jet \pT\ region ($y$-axis) is shown.
The leading jet's \pT\ in the SRL is denoted by $p_{\text{T}}^{\text{L}}$ and the subleading jet's \pT\ in the SRS is denoted by $p_{\text{T}}^{\text{S}}$.
\label{fig:STXSNumbers}}
\end{figure}

\begin{table}[tb]
\begin{center}
\caption{Signal acceptance times efficiency for the STXS volumes in the differential measurement.
Along with the \pTH\ requirements shown, $\absyH<2$ is required.
For events with $\pTH<300~\GeV$, the acceptance times efficiency is less than \num{0.1e-2}.
\label{tab:STXSAccept}}
\begin{tabular}{lcccc}
\toprule
Process & $300<\pTH<450~\GeV$ & $450<\pTH<650~\GeV$  & $650<\pTH<1000~\GeV$ & $\pTH>1~\TeV$ \\
\midrule
All & \num{1.3e-2} & 0.23 & 0.31 & 0.23 \\
\midrule
\ggF & \num{0.7e-2} & 0.25 & 0.35 & 0.28 \\
\VBF & \num{0.4e-2} & 0.21 & 0.32 & 0.25 \\
\VH  & \num{1.7e-2} & 0.26 & 0.30 & 0.20 \\
\ttH & \num{4.7e-2} & 0.19 & 0.24 & 0.19 \\
\bottomrule
\end{tabular}
\end{center}
\end{table}


Again, the procedure is tested with \Wqq and \Zqq in the VR and \Zbb\ in the SR
with the $V$ and $Z$ mass templates structured similarly to those of the Higgs boson described above.
Due to the larger sample size, the VRL is divided into five slices, the fit is performed independently on each slice, and the results are averaged. In the SR, the $Z$ fit is performed in the SRL, SRS, and \CRtt regions with the Higgs boson contribution fixed to the SM prediction. Again, the results do not change when using a freely floating Higgs boson normalization. Results of the two differential fits are shown in Figure~\ref{fig:kFSTXS}, where they are compared with the predictions for the EW NLO and QCD NNLO corrections as a function of reconstructed \pTV.
Both results agree with SM expectations.
 
\begin{figure}[h]
\begin{center}
\begin{subfigure}{0.45\textwidth}
\centering
\includegraphics[width=\textwidth]{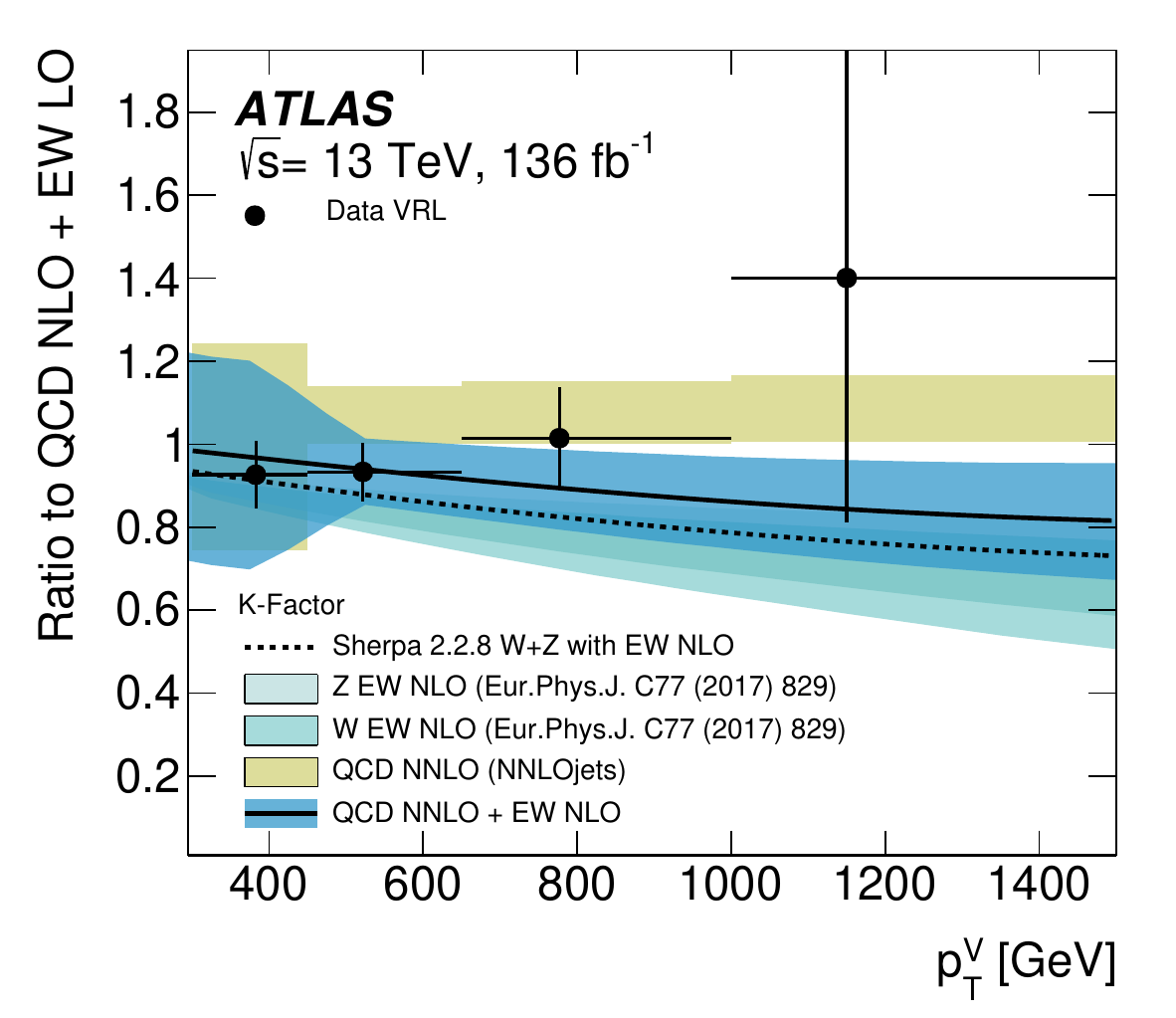}
\caption{\Vjets in VRL}
\end{subfigure}
\begin{subfigure}{0.45\textwidth}
\centering
\includegraphics[width=\textwidth]{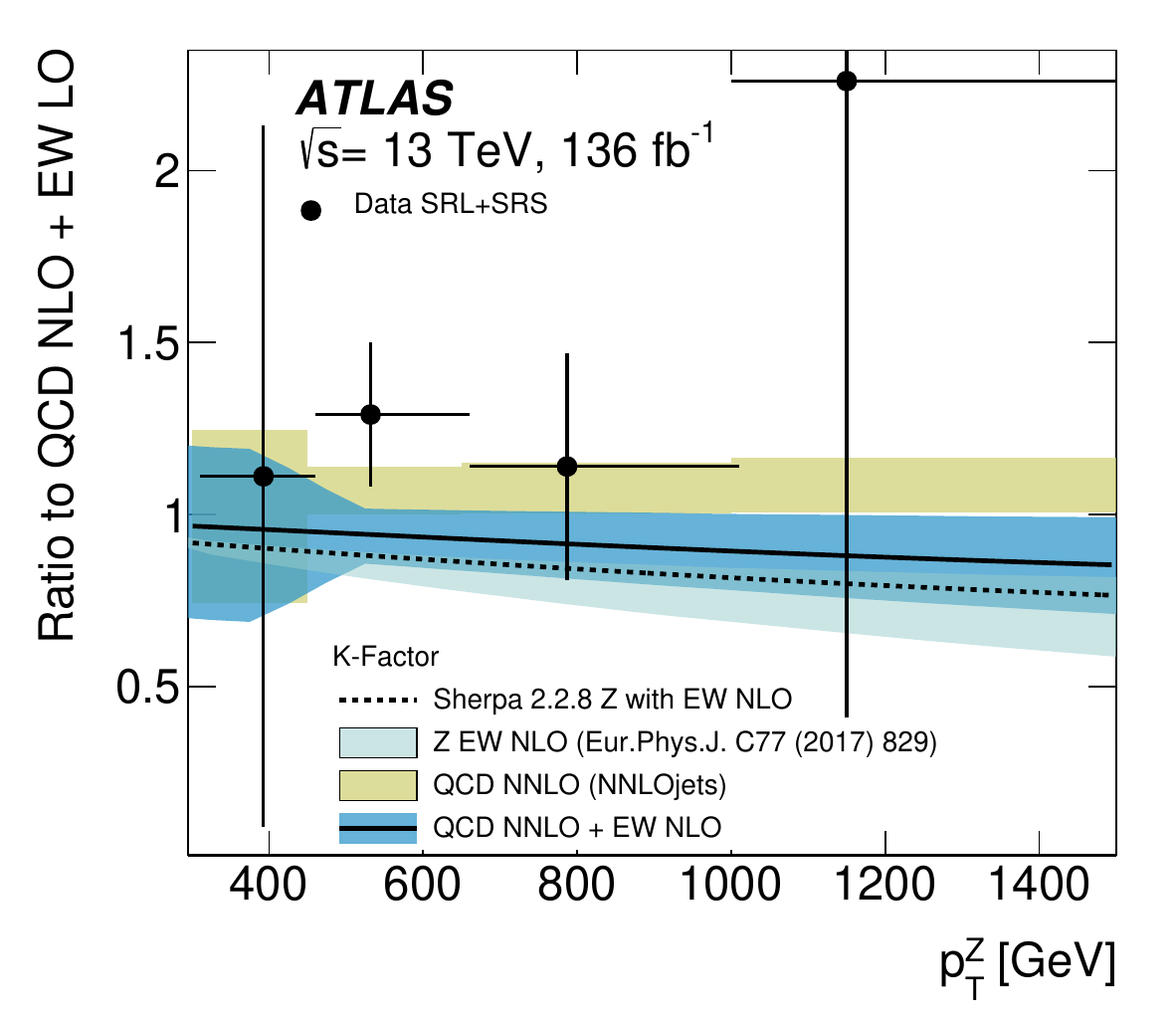}
\caption{\Zjets in SR}
\end{subfigure}
\caption{Comparison of differential fit signal strengths for (a) \Vjets in the VRL and (b) \Zjets in the SR.
The signal strength within the STXS volumes is calculated relative to the prediction at NLO QCD and LO EW accuracy.
They are compared with the NLO EW correction provided by \SHERPA, the NNLO QCD correction provided by the NNLOJET group, and their product.
The Higgs boson yields are kept fixed to the SM expectation when extracting the \Zjets\ signal strength within the STXS volumes.
The points are located at the weighted center of the bin considering the underlying \pTV\ or \pTH\ spectrum
\label{fig:kFSTXS}}
\end{center}
\end{figure}

To extract the four Higgs boson signal strengths within the STXS volumes, ten differential SR and CR regions defined in Table~\ref{tab:SRs} are fitted simultaneously, exploiting the corresponding systematic uncertainty configurations shown in Table~\ref{tab:systs}. The results are summarized in Table~\ref{tab:STXSData} and Figure~\ref{fig:STXSCorr}.
The \ttbar normalization corrections determined from the data in each jet \pT\ region are compatible with \ttbar event measurements in a similar kinematic phase-space~\cite{TOPQ-2018-15}.
The four Higgs boson signal strengths are compatible, with a $p$-value of 0.53.
Post-fit jet mass distributions from the SRL are shown in Figure~\ref{fig:SRLSTXSPostFit}.

\begin{table}[tb]
\caption{Expected and observed values of the signal strengths for the \Hboson, \Zboson, and \ttbar\ components in the differential fits. The value for \ensuremath{\mu_{H}}\xspace refers to a STXS volume, while those for \ensuremath{\mu_{Z}}\xspace and \ensuremath{\mu_{\ttbar}}\xspace pertain to the corresponding jet \pT regions.
Absolute cross section uncertainties are not included in the reported \ensuremath{\mu_{Z}}\xspace and \ensuremath{\mu_{\ttbar}}\xspace values.
\label{tab:STXSData}}
\footnotesize
\begin{center}
\begin{tabular}{c
S[table-format= 1.1(2)] 
S[table-format=-1.0(2)] 
}
\toprule
\multirow{2}{*}{\pTH[GeV]}  & \multicolumn{2}{c}{\ensuremath{\mu_{H}}\xspace} \\
& {Exp.} & {Obs.} \\
\midrule
300--450  & 1 \pm 18   &  -6   \pm 18  \\
450--650  & 1 \pm  3.3 &  -3   \pm 5   \\
~~650--1000 & 1 \pm  6   &   5   \pm 7   \\
$>$1000   & 1 \pm 30   &  18   \pm 32  \\
\bottomrule
\end{tabular}
\quad
\quad
\begin{tabular}{c
S[table-format= 1.2(2)] 
S[table-format= 1.2(2)] 
S[table-format= 1.2(2)] 
S[table-format= 1.2(2)] 
}
\toprule
\multirow{2}{*}{Jet \pT[GeV]}  &  \multicolumn{2}{c}{\ensuremath{\mu_{Z}}\xspace} & \multicolumn{2}{c}{\ensuremath{\mu_{\ttbar}}\xspace} \\
& {Exp.} & {Obs.} & {Exp.} & {Obs.} \\
\midrule
250--450 & 1.0\pm1.1   &  1.8  \pm 1.1  & 1.0\pm0.07  &  0.85 \pm 0.06 \\
450--650 & 1.0\pm0.17  &  1.28 \pm 0.22 & 1.0\pm0.07  &  0.76 \pm 0.06 \\
~~650--1000& 1.0\pm0.33  &  1.4  \pm 0.4  & 1.0\pm0.09  &  0.74 \pm 0.08 \\
$>$1000  & 1.0\pm1.6   &  2.4  \pm 1.7  & 1.0\pm0.22  &  0.57 \pm 0.18 \\
\bottomrule
\end{tabular}
\end{center}
\end{table}


\begin{figure}[h]
\begin{center}
\begin{subfigure}{0.49\textwidth}
\centering
\includegraphics[width=\textwidth]{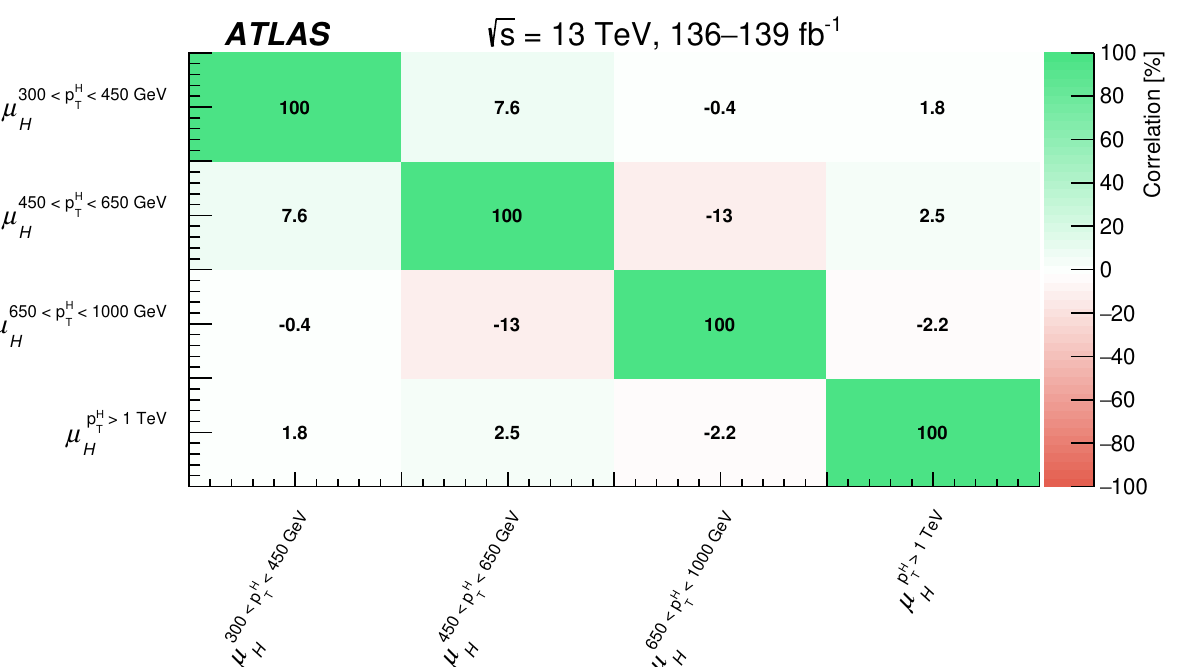}
\end{subfigure}
\begin{subfigure}{0.49\textwidth}
\centering
\includegraphics[width=\textwidth]{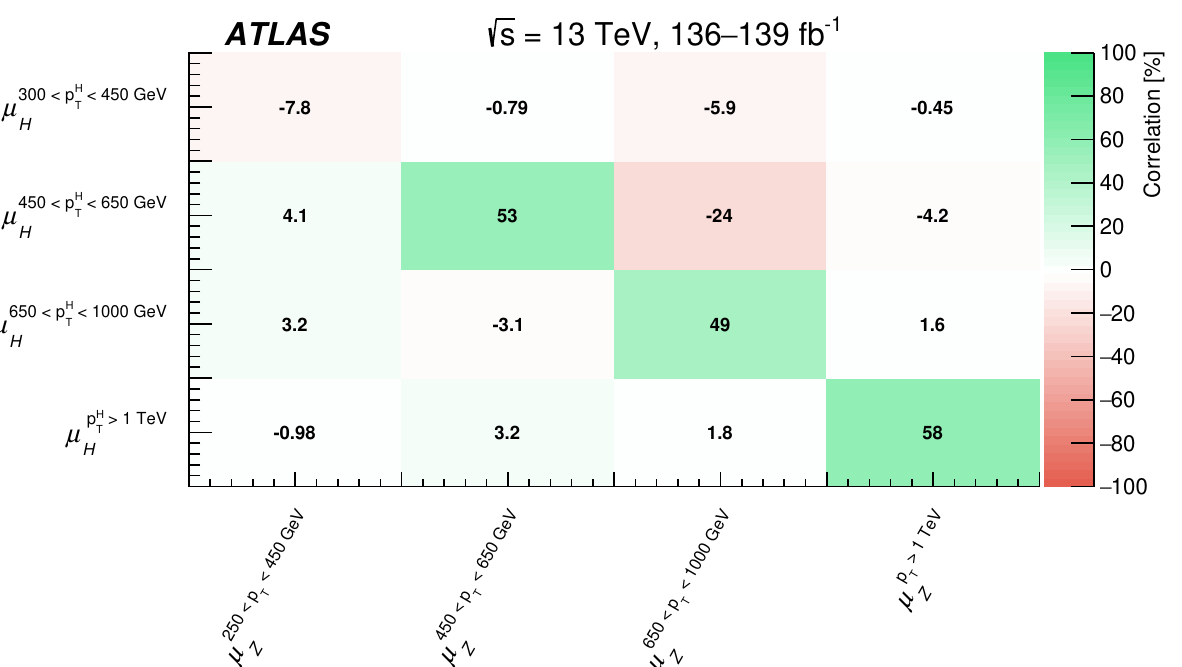}
\end{subfigure}
\end{center}
\caption{
Correlations among the four Higgs boson signal strengths, and
between the four Higgs boson and \Zjets\ signal strengths.
The Higgs boson signal strengths \ensuremath{\mu_{H}}\xspace are labeled with the corresponding \pTH range as a superscript.
The \Zjets\ signal strengths \ensuremath{\mu_{Z}}\xspace are labeled with the corresponding jet \pT range as a superscript.
\label{fig:STXSCorr}}
\end{figure}

\def\pTRegCaption{The five STXS volumes are labeled \pTReg{0}--\pTReg{4} corresponding to $\pTH<300~\GeV$, 300--450~\GeV, 450--650~\GeV, 650--1000~\GeV, and ${>}1000~\GeV$, respectively. The \pTReg{0} event yield is constrained to its SM value within the theoretical and experimental uncertainties and free parameters act independently on the remaining four volumes. }
 
\begin{figure}[htb]
\begin{center}
\begin{subfigure}{0.30\textwidth}
\centering
\includegraphics[width=\textwidth]{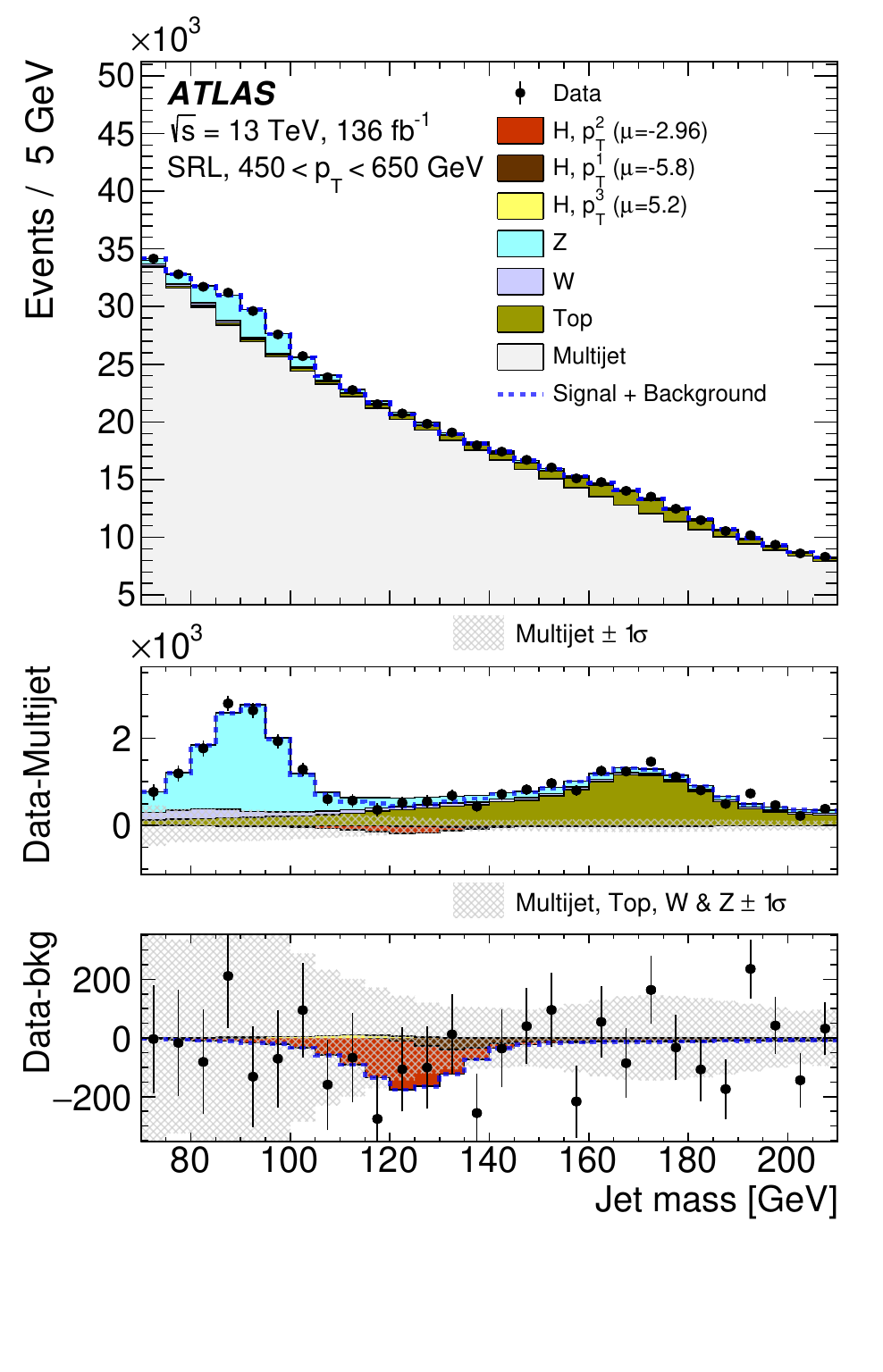}
\end{subfigure}
\begin{subfigure}{0.30\textwidth}
\centering
\includegraphics[width=\textwidth]{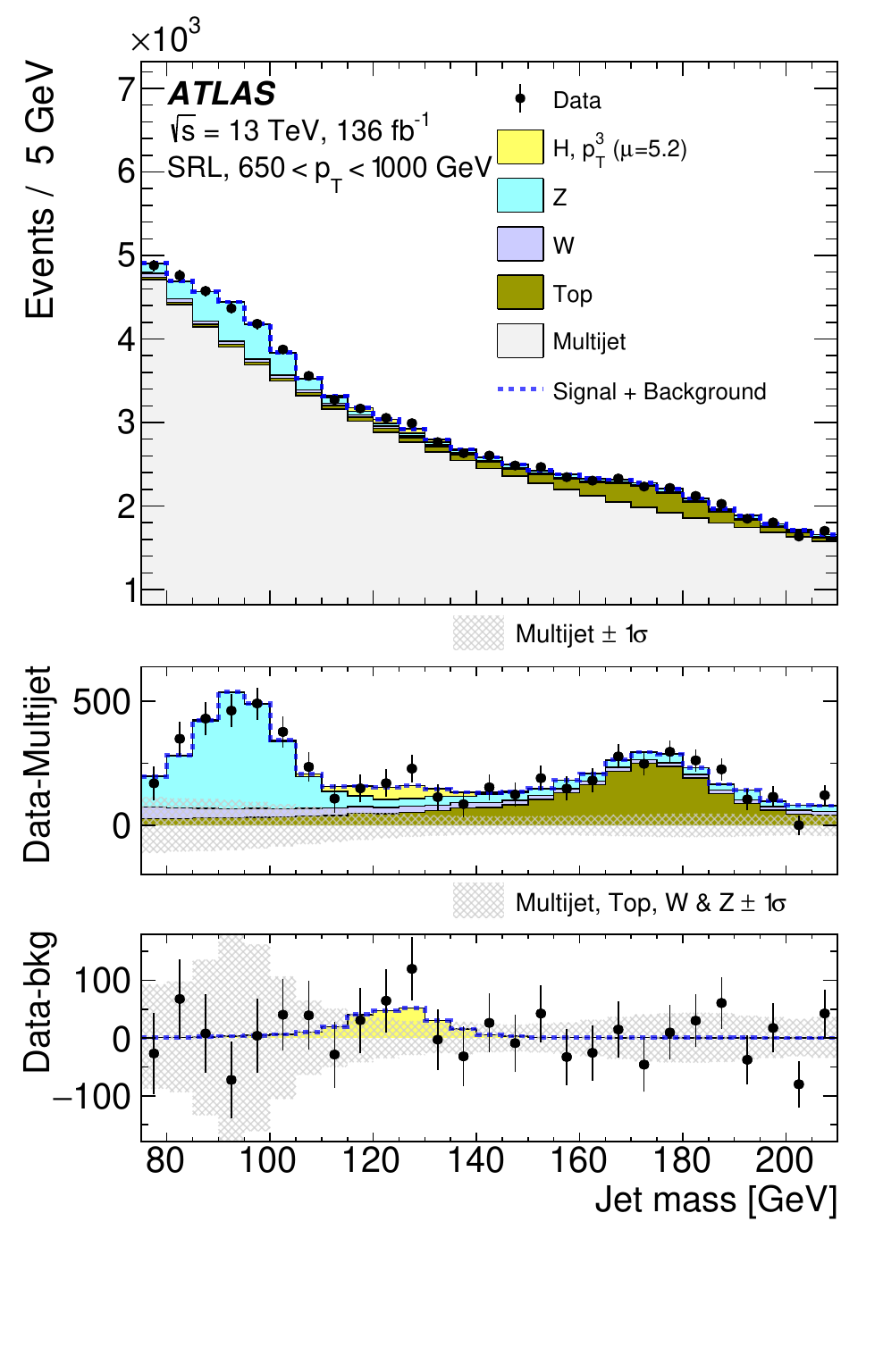}
\end{subfigure}
\begin{subfigure}{0.30\textwidth}
\centering
\includegraphics[width=\textwidth]{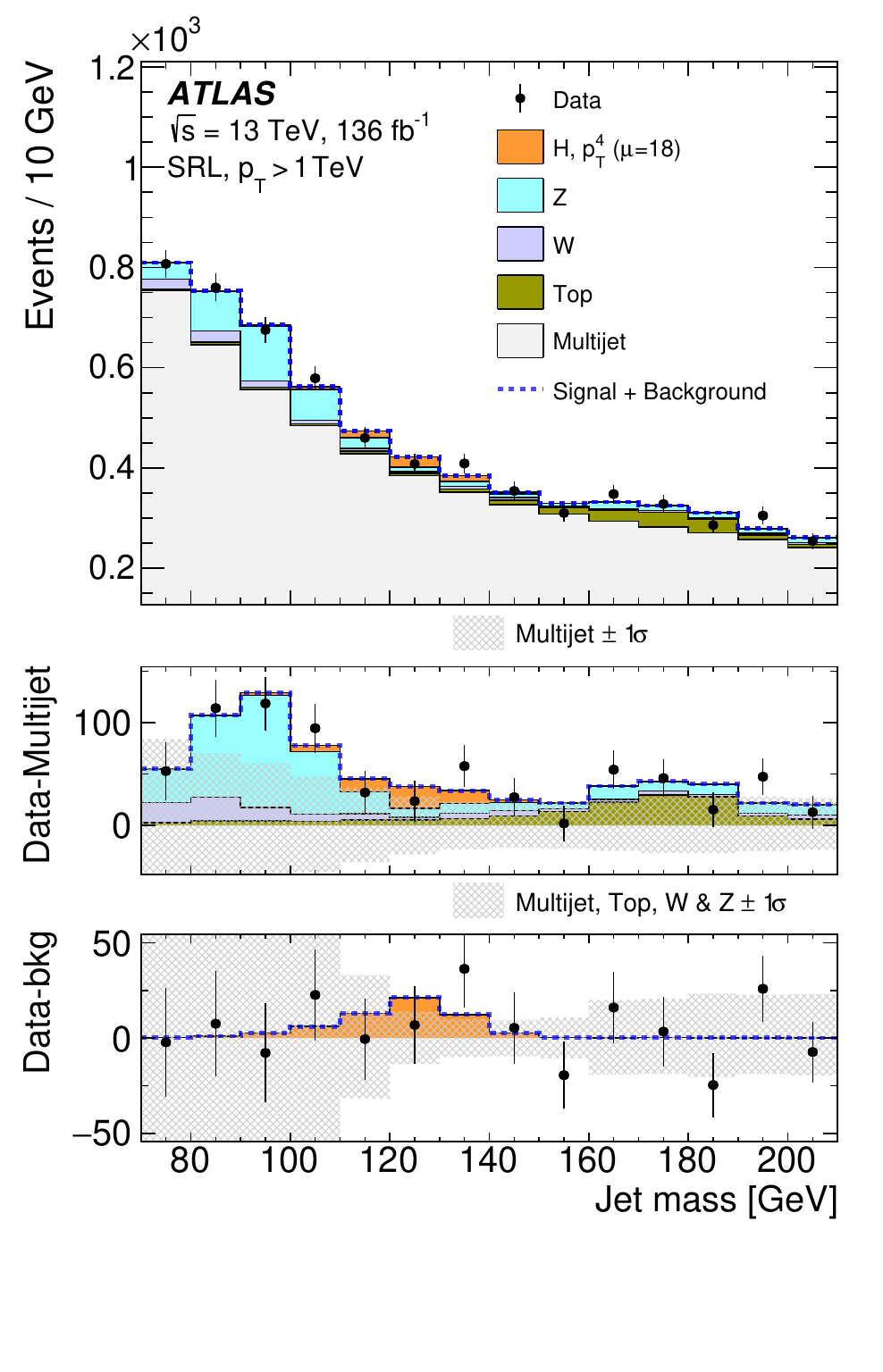}
\end{subfigure}
\end{center}
\caption{Post-fit jet mass distributions of the various components in the differential leading-jet signal region defined by the selected candidate jet with $450<\pT<650~\GeV$ (left), $650<\pT<1000~\GeV$ (middle), and $\pT>1000~\GeV$ (right) shown in wider 10~\GeV\ bins.
\SRPostFitCapMid
\SRPostFitCapLower
\pTRegCaption
Contributions below 0.5~per~mille of the total yield are not shown.
\label{fig:SRLSTXSPostFit}}
\end{figure}

The resulting Higgs boson production cross section for $\pTH>1~\TeV$ is
\begin{equation*}
\sigma_{H}(\pTH>1~\TeV) = 2.3 \pm 3.9~\mathrm{(stat.)} \pm 1.3~\mathrm{(syst.)} \pm 0.5~\mathrm{(theory)}~\mathrm{fb}
\end{equation*}
and the SM prediction is 0.13~fb.
The differential results correspond to the following observed (expected) 95\% CL upper limits on the Higgs boson differential production cross sections:
\begin{eqnarray*}
\sigma_{H}(300<\pTH<450~\GeV)  &<& 2.9~(3.1) \mathrm{~pb,} \\
\sigma_{H}(450<\pTH<650~\GeV)  &<& 89 ~(102) \mathrm{~fb,} \\
\sigma_{H}(650<\pTH<1000~\GeV) &<& 39 ~(34)  \mathrm{~fb,} \\
\sigma_{H}(\pTH>1000~\GeV)     &<& 9.6~(7.4) \mathrm{~fb.}
\end{eqnarray*}
 
The fitted values and upper limits of the Higgs boson signal strengths in four \pTH\ intervals are shown in Figure~\ref{fig:summary}.
The largest source of uncertainty is the data sample size. The contributions from the main sources of uncertainty are summarized in Table~\ref{tab:STXSSysts}.
The jet uncertainties tend to give the largest contribution of systematic uncertainty, driven by JMS effects.
The dominant JMS uncertainty component in \Vjets\ events is reduced by 30\%--40\% separately in the three SRs with jet $\pT>450~\GeV$. The same component is reduced by 20\%--65\% for \T\ events over the full jet \pT range.
The JMS nuisance parameter receives small pulls that differ between \Vjets\ and \T\ events and with \pT.
The impact of JMR increases above 1~\TeV, where the JMR measurement constraints on \Vjets are looser due to the extrapolation uncertainties (see Figure~\ref{fig:WMassResVsPt}). The JMR uncertainty for jets in \T\ events is constrained to a relative 6\%--9\% (from 20\%) below 1~\TeV\ but is 18\% above 1~\TeV.
Similarly, the flavor-tagging uncertainties increase above 1~\TeV\ due to the extrapolation from the \pT range of the calibration regions.
The \ttbar modeling systematic uncertainties are more relevant in the first jet-\pT bin and decrease above $450~\GeV$, where the top quark decay products become more collimated, thus reducing the contamination around the Higgs boson mass peak.
The fit reduces these systematic uncertainties by 20\%--30\%, driven by the purity of the \CRtt.
Other than the few mentioned, no other nuisance parameter has a posterior probability very different from the prior.

\begin{figure}[htpb!]
\centering
\includegraphics[width=0.60\textwidth]{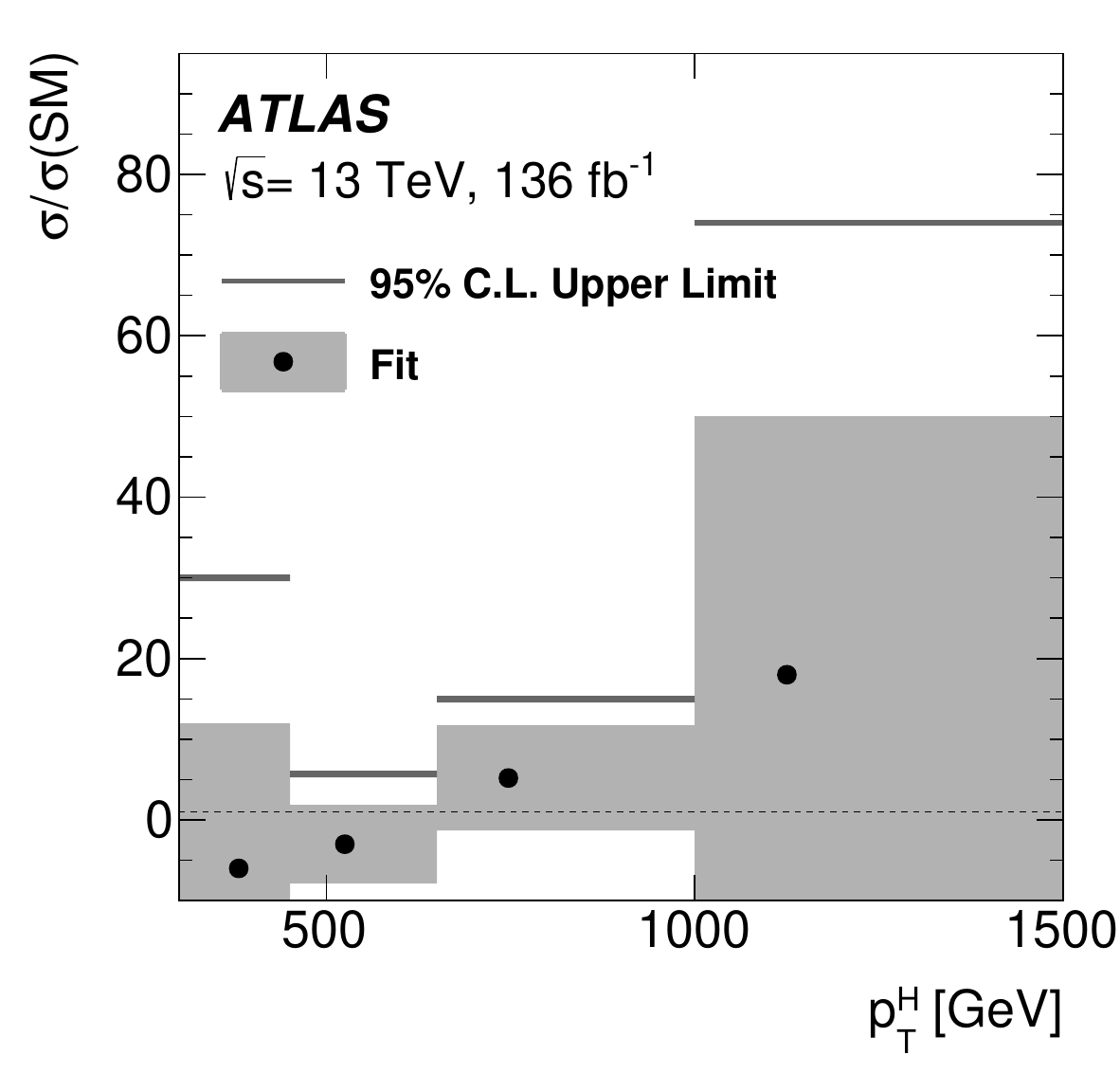}
\caption{
Summary of the STXS volume signal strengths measured using the differential signal regions.
Within the same kinematic regimes, measurements of the $Z\rightarrow b\bar{b}$ process agree with the Standard Model predictions, validating the methods.
The points are located at the weighted center of the bin considering the underlying \pTH\ spectrum.
\label{fig:summary}}
\end{figure}

\begin{table}[tb]
\caption{Contributions to the systematic uncertainties for the differential measurements of the signal-strength, defined as the signal yield relative to the SM prediction within STXS volumes. The total uncertainty is also given for comparison.
\label{tab:STXSSysts}}
\footnotesize
\begin{center}
\begin{tabular}{l S[table-format=<2.1] S[table-format=<2.1] S[table-format=<2.1] S[table-format=2.1]}
\toprule
Uncertainty Contribution & {$300< \pTH < 450$~GeV} & {$450< \pTH < 650$~GeV} & {$650< \pTH < 1000$~GeV} & {$\pTH > 1$~TeV} \\
\midrule
Total                       & 18   &  5.0 &  6.5  & 32   \\
\midrule
Statistical                 & 16   &  3.0 &  5.5  & 30   \\
Systematic                  &  7   &  3.9 &  3.4  & 10   \\
\midrule
Jet systematic uncertainties             &  6   &  3.8 &  3.4  &  9.5 \\
Modeling and theory systs.  &  4   &  0.7 &  0.7  &  2   \\
Flavor-tagging systs.       &  0.2 &  0.4 &  0.4  &  2   \\
\bottomrule
\end{tabular}
\end{center}
\end{table}



\FloatBarrier


\label{sec:results}
\section{Conclusions}

High-\pT Higgs boson production is studied in the $b \bar{b}$ decay channel probing a new highly Lorentz-boosted boson reconstruction domain with transverse momentum above 1 TeV.
The results are based on $pp$ collision data collected at $\sqrt{s}=13$~\TeV\ with the ATLAS detector during Run~2 of the LHC, corresponding to an integrated luminosity of 136~\ifb.
The Higgs boson is reconstructed as a single large-$R$ jet and identified with $b$-tagging techniques.
The measured signal strengths of \Wqq\ and \Zqq\ in the validation region and \Zbb\ in the signal region agree with the SM predictions and validate the experimental techniques.

The observed (expected) 95\% CL limit on the Higgs boson production cross section for $\pTH>450$~\GeV\ obtained from the fiducial signal region is 115 (128)~fb.
From the four differential signal regions, the observed (expected) 95\% CL limits on the Higgs boson production cross section are:
\begin{eqnarray*}
\sigma_{H}(300<\pTH<450~\GeV)  &<& 2.9 (3.1) \mathrm{~pb,} \\
\sigma_{H}(450<\pTH<650~\GeV)  &<& 89  (102) \mathrm{~fb,} \\
\sigma_{H}(650<\pTH<1000~\GeV) &<& 39  (34)  \mathrm{~fb,} \\
\sigma_{H}(\pTH>1000~\GeV)     &<& 9.6 (7.4) \mathrm{~fb.}
\end{eqnarray*}
The Higgs boson production cross section for $\pTH>1~\TeV$ is found to be
\begin{equation*}
\sigma_{H}(\pTH>1~\TeV) = 2.3 \pm 3.9~\mathrm{(stat.)} \pm 1.3~\mathrm{(syst.)} \pm 0.5~\mathrm{(theory)}~\mathrm{fb.} \\
\end{equation*}
All of the Higgs boson results are consistent with the Standard Model predictions.


\section*{Acknowledgments}


We thank CERN for the very successful operation of the LHC, as well as the
support staff from our institutions without whom ATLAS could not be
operated efficiently.

We acknowledge the support of 
ANPCyT, Argentina; 
YerPhI, Armenia; 
ARC, Australia; 
BMWFW and FWF, Austria; 
ANAS, Azerbaijan; 
SSTC, Belarus; 
CNPq and FAPESP, Brazil; 
NSERC, NRC and CFI, Canada; 
CERN; 
ANID, Chile; 
CAS, MOST and NSFC, China; 
Minciencias, Colombia; 
MEYS CR, Czech Republic; 
DNRF and DNSRC, Denmark; 
IN2P3-CNRS and CEA-DRF/IRFU, France; 
SRNSFG, Georgia; 
BMBF, HGF and MPG, Germany; 
GSRI, Greece; 
RGC and Hong Kong SAR, China; 
ISF and Benoziyo Center, Israel; 
INFN, Italy; 
MEXT and JSPS, Japan; 
CNRST, Morocco; 
NWO, Netherlands; 
RCN, Norway; 
MEiN, Poland; 
FCT, Portugal; 
MNE/IFA, Romania; 
JINR; 
MES of Russia and NRC KI, Russian Federation; 
MESTD, Serbia; 
MSSR, Slovakia; 
ARRS and MIZ\v{S}, Slovenia; 
DSI/NRF, South Africa; 
MICINN, Spain; 
SRC and Wallenberg Foundation, Sweden; 
SERI, SNSF and Cantons of Bern and Geneva, Switzerland; 
MOST, Taiwan; 
TAEK, Turkey; 
STFC, United Kingdom; 
DOE and NSF, United States of America. 
In addition, individual groups and members have received support from 
BCKDF, CANARIE, Compute Canada and CRC, Canada; 
COST, ERC, ERDF, Horizon 2020 and Marie Sk{\l}odowska-Curie Actions, European Union; 
Investissements d'Avenir Labex, Investissements d'Avenir Idex and ANR, France; 
DFG and AvH Foundation, Germany; 
Herakleitos, Thales and Aristeia programmes co-financed by EU-ESF and the Greek NSRF, Greece; 
BSF-NSF and GIF, Israel; 
Norwegian Financial Mechanism 2014-2021, Norway; 
NCN and NAWA, Poland; 
La Caixa Banking Foundation, CERCA Programme Generalitat de Catalunya and PROMETEO and GenT Programmes Generalitat Valenciana, Spain; 
G\"{o}ran Gustafssons Stiftelse, Sweden; 
The Royal Society and Leverhulme Trust, United Kingdom.

The crucial computing support from all WLCG partners is acknowledged gratefully, in particular from CERN, the ATLAS Tier-1 facilities at TRIUMF (Canada), NDGF (Denmark, Norway, Sweden), CC-IN2P3 (France), KIT/GridKA (Germany), INFN-CNAF (Italy), NL-T1 (Netherlands), PIC (Spain), ASGC (Taiwan), RAL (UK) and BNL (USA), the Tier-2 facilities worldwide and large non-WLCG resource providers. Major contributors of computing resources are listed in Ref.~\cite{ATL-SOFT-PUB-2023-001}.


\clearpage
\printbibliography
 
\clearpage

\clearpage
 
\begin{flushleft}
\hypersetup{urlcolor=black}
{\Large The ATLAS Collaboration}

\bigskip

\AtlasOrcid[0000-0002-6665-4934]{G.~Aad}$^\textrm{\scriptsize 99}$,
\AtlasOrcid[0000-0002-5888-2734]{B.~Abbott}$^\textrm{\scriptsize 117}$,
\AtlasOrcid[0000-0002-7248-3203]{D.C.~Abbott}$^\textrm{\scriptsize 100}$,
\AtlasOrcid[0000-0002-2788-3822]{A.~Abed~Abud}$^\textrm{\scriptsize 34}$,
\AtlasOrcid[0000-0002-1002-1652]{K.~Abeling}$^\textrm{\scriptsize 53}$,
\AtlasOrcid[0000-0002-2987-4006]{D.K.~Abhayasinghe}$^\textrm{\scriptsize 92}$,
\AtlasOrcid[0000-0002-8496-9294]{S.H.~Abidi}$^\textrm{\scriptsize 27}$,
\AtlasOrcid[0000-0002-9987-2292]{A.~Aboulhorma}$^\textrm{\scriptsize 33e}$,
\AtlasOrcid[0000-0001-5329-6640]{H.~Abramowicz}$^\textrm{\scriptsize 148}$,
\AtlasOrcid[0000-0002-1599-2896]{H.~Abreu}$^\textrm{\scriptsize 147}$
% ...
\bigskip
\\

% ...
$^{27}$Physics Department, Brookhaven National Laboratory, Upton NY; United States of America.\\
% ...
$^{33}$$^{(a)}$Facult\'e des Sciences Ain Chock, R\'eseau Universitaire de Physique des Hautes Energies - Universit\'e Hassan II, Casablanca;$^{(b)}$Facult\'{e} des Sciences, Universit\'{e} Ibn-Tofail, K\'{e}nitra;$^{(c)}$Facult\'e des Sciences Semlalia, Universit\'e Cadi Ayyad, LPHEA-Marrakech;$^{(d)}$LPMR, Facult\'e des Sciences, Universit\'e Mohamed Premier, Oujda;$^{(e)}$Facult\'e des sciences, Universit\'e Mohammed V, Rabat;$^{(f)}$Institute of Applied Physics, Mohammed VI Polytechnic University, Ben Guerir; Morocco.\\
$^{34}$CERN, Geneva; Switzerland.\\
% ...
$^{53}$II. Physikalisches Institut, Georg-August-Universit\"{a}t G\"ottingen, G\"ottingen; Germany.\\
% ...
$^{92}$Department of Physics, Royal Holloway University of London, Egham; United Kingdom.\\
% ...
$^{99}$CPPM, Aix-Marseille Universit\'e, CNRS/IN2P3, Marseille; France.\\
$^{100}$Department of Physics, University of Massachusetts, Amherst MA; United States of America.\\
% ...
$^{117}$Homer L. Dodge Department of Physics and Astronomy, University of Oklahoma, Norman OK; United States of America.\\
% ...
$^{147}$Department of Physics, Technion, Israel Institute of Technology, Haifa; Israel.\\
$^{148}$Raymond and Beverly Sackler School of Physics and Astronomy, Tel Aviv University, Tel Aviv; Israel.\\
% ...
$^{a}$ Also Affiliated with an institute covered by a cooperation agreement with CERN.\\
$^{b}$ Also at Borough of Manhattan Community College, City University of New York, New York NY; United States of America.\\
$^{c}$ Also at Bruno Kessler Foundation, Trento; Italy.\\
$^{d}$ Also at Center for High Energy Physics, Peking University; China.\\
$^{e}$ Also at Centro Studi e Ricerche Enrico Fermi; Italy.\\
$^{f}$ Also at CERN, Geneva; Switzerland.\\
% ...

\end{flushleft}


\end{document}